\documentclass[onecolumn,11pt,final,dvips]{report}

\usepackage{times}
\usepackage{amsmath,dsfont}
\usepackage{amssymb,amsthm}
\usepackage{epsfig,verbatim}
\usepackage{subfigure}
\usepackage{setspace}
\usepackage{color}
\usepackage{graphicx}
\usepackage[top=1in, bottom=1.25in, left=1.25in, right=1.25in]{geometry}
\definecolor{gray}{rgb}{0.7,0.7,0.7}
\definecolor{lgray}{rgb}{0.75,0.87,0.87}
\definecolor{dred}{rgb}{0.6, 0, 0}
\definecolor{ddred}{rgb}{0.4, 0, 0}
\definecolor{orange}{cmyk}{0,0.5,1,0.1}
\definecolor{dorange}{cmyk}{0,0.6,1,0.4}
\definecolor{dblue}{rgb}{0,0,0.5}
\definecolor{ddblue}{rgb}{0,0,0.2}
\definecolor{lblue}{rgb}{.5,.5,1}
\definecolor{dgreen}{rgb}{0,0.28,0}
\definecolor{dgreen2}{rgb}{0,0.5,0.2}
\definecolor{dgreen3}{rgb}{0,0.6,0.3}
\definecolor{mgreen}{rgb}{0,0.6,0}
\definecolor{dgray}{rgb}{0.4,0.4,0.4}
\definecolor{dbrown}{rgb}{0.62,0.33,0.18}
\definecolor{brown}{cmyk}{0.5,0.7,0.7,0}
\definecolor{white}{rgb}{1,1,1}
\definecolor{lred}{rgb}{1,.7,.7}
\definecolor{dpurple}{rgb}{0.3,0,0.9}
\definecolor{purple}{rgb}{0.5,0,0.5}
\definecolor{lpurple}{rgb}{0.3,0,0.3}
\definecolor{dpink}{cmyk}{.2,1,.7,0}

\newtheorem{definition}{Definition}
\newtheorem{theorem}{Theorem}
\newtheorem{corollary}{Corollary}
\newtheorem{proposition}{Proposition}
\newtheorem{lemma}{Lemma}

\newcommand{\E}{\mathds{E}}

\long\def\symbolfootnote[#1]#2{\begingroup\def\thefootnote{\fnsymbol{footnote}}\footnote[#1]{#2}\endgroup}

\title{\vspace{-1.5cm}\doublespacing
\fontsize{25pt}{10}\selectfont {\bf On the Capacity of Narrowband PLC Channels} \\ \vspace{1.0cm}
\fontsize{15pt}{10}\selectfont Nir Shlezinger and Ron Dabora\:  \\ \vspace{0.7cm}
\fontsize{15pt}{10}\selectfont {\bf Email: nirshl@post.bgu.ac.il, ron@ee.bgu.ac.il}
}
\date{\doublespacing
\bigskip
\bigskip
\vspace{2cm}
\fontsize{15pt}{10}\selectfont A complete version with detailed proof of the paper which was published in the IEEE Transactions on Communications, Vol. 63, No. 4, Apr. 2015, pp. 1191-1201. \\
\vspace{5cm}
\fontsize{15pt}{10}\selectfont
This work was supported by the Israeli Ministry of Economy through the Israeli Smart Grid Consortium (ISG) }
\author{
}

\begin{document}

\maketitle
\thispagestyle{empty}
\cleardoublepage
\chapter*{Abstract}
\label{chap_star:Background}
		Power line communications (PLC) is the central communications technology for the realization of smart power grids.
		As the designated band for smart grid communications is the narrowband (NB) power line channel, NB-PLC  has been receiving substantial attention in recent years. Narrowband power line channels are characterized by cyclic short-term variations of the channel transfer function (CTF) and strong noise with periodic statistics.
		In this work, modeling the CTF as a linear periodically time-varying filter and the noise as an additive cyclostationary Gaussian process,
			we derive the capacity of discrete-time NB-PLC channels.
		As part of the capacity derivation, we characterize the capacity achieving transmission scheme, which leads to a practical code construction that approaches capacity.
		The capacity derived in this work is numerically evaluated for several NB-PLC channel
configurations taken from previous works, and the results show that the optimal scheme achieves a substantial rate gain over a previously proposed ad-hoc scheme. This gain is due to optimally accounting for the periodic properties of the channel and the noise.
\chapter*{Keywords}
\label{chap_star:keywords}

Power line communications, cyclostationary noise, channel capacity.

\pagenumbering{roman}
%
\tableofcontents
\listoffigures

\newpage
\pagenumbering{arabic}

\vspace{-0.45cm}
\chapter{Introduction}
\label{sec:Intro}
\vspace{-0.1cm}
The frequency band designated for the automation and control of power line networks is the frequency range of $3 - 500$ kHz  \cite{Galli:11}, \cite{Ferreira:10}. Communications over this band is referred to as narrowband (NB) power line communications (PLC), and is an essential component in the realization of smart power grids. As the transition to smart power grids accelerates, the communications requirements for network control increase accordingly. Therefore, characterizing the fundamental rate limits for NB-PLC channels and establishing guidelines for the optimal transmission scheme are essential to the successful implementation of future smart power grids.

Unlike conventional wired communications media, the power line was not designed for bi-directional communications, but for uni-directional power transfer, and its characteristics are considerably affected by electrical appliances connected to the power grid \cite{Ferreira:10}.
It has been established that the narrowband power line channel exhibits multipath signal propagation \cite{Zimmermann:02a, Prassana:09}, and that the channel transfer function (CTF)  and the noise statistics in narrowband power line channels vary periodically with respect to the mains frequency \cite{Dostert:11, Evans:12, Katayama:08}.
In accordance, the NB-PLC CTF is commonly modeled as a passband linear periodically time-varying (LPTV) filter \cite{Dostert:11, Evans:12, Katayama:08},
and the NB-PLC noise is modeled as 
a passband additive {\em cyclostationary} Gaussian noise (ACGN) \cite{Ferreira:10, Evans:12, Katayama:06, Nassar:12}.
The overall NB-PLC channel is thus modeled as a passband LPTV channel with ACGN.

The capacity of NB-PLC channels was considered previously in \cite{Hooijen:98a}. However, \cite{Hooijen:98a} assumed that the CTF is {\em time-invariant} and that the noise is {\em stationary}.
As the  NB-PLC channel has fundamentally different characteristics, the capacity expression of \cite{Hooijen:98a} does not represent the actual capacity of NB-PLC channels.
In \cite{Han:12} the capacity of continuous time (CT) linear time-invariant (LTI) channels with ACGN was investigated via the harmonic series representation (HSR).
%
%
To the best of our knowledge, {\em no previous work derived the capacity of discrete-time (DT) NB-PLC channels, considering the periodic properties of both the channel and the noise}.
In this work we fill this gap.

Lastly, we note that the work  \cite{Katayama:09} proposed a practical transmission scheme, based on orthogonal frequency division multiplexing (OFDM), for LTI channels with ACGN , and \cite{Lampe:13} proposed a practical OFDM scheme for LPTV channels with AWGN.
Following the same design principles as in  \cite{Katayama:09} and \cite{Lampe:13}, we can obtain a practical transmission scheme, based on OFDM, which accounts for the periodic properties of both the CTF and the noise statistics. We will show that the optimal scheme derived in this work is substantially different from the practical scheme which follows the considerations used in previous works, and that the optimal scheme achieves significantly better performance.
%

\vspace{-0.25cm}
\section *{Main Contributions and Organization}
In this paper, we derive the capacity of DT NB-PLC channels, accounting for the periodic nature of the CTF and the periodic statistics of the noise by modeling the CTF as an LPTV filter \cite{Dostert:11, Evans:12, Katayama:08}, and modeling the noise as an ACGN \cite{Katayama:06, Nassar:12}.
We present two capacity derivations: Our first derivation applies the decimated components decomposition (DCD) to transform the channel into a static multiple-input multiple-output (MIMO) channel with no intersymbol interference (ISI) and with multivariate additive white Gaussian noise (AWGN), and then obtains the NB-PLC capacity as the capacity of the asymptotic MIMO model, obtained by taking the dimensions to infinity. Our second derivation uses the DCD to transform the channel into an LTI MIMO channel with finite memory and with additive colored Gaussian noise, and then obtains the NB-PLC capacity from the capacity of finite memory Gaussian MIMO channels derived in \cite{Brandenburg:74}.
Our work is fundamentally different from \cite{Han:12} in two major aspects:
First, note that the capacity analysis of DT channels is fundamentally different from the capacity analysis of CT channels \cite[Ch. 9.3]{Cover:00}. Second, applying the approach of \cite{Han:12}, namely the HSR, to our DT NB-PLC scenario leads to a transformed DT LTI MIMO channel with {\em infinite} ISI and with AWGN, {\em whose capacity is unknown}.
We conclude that applying the DCD is the most appropriate approach for analyzing the capacity of DT LPTV channels with ACGN.
From the numerical results it follows that in the practical scenarios studied, the capacity of NB-PLC channels is substantially higher than the achievable rate of the practical scheme based on \cite{Katayama:09} and \cite{Lampe:13}.

The rest of this paper is organized as follows: In Chapter \ref{sec:Preliminaries} the relevant properties of cyclostationary processes are briefly recalled and the DCD is presented.
In Chapter \ref{sec:Model} the channel model is presented and a practical scheme based on \cite{Katayama:09} and \cite{Lampe:13} is reviewed, in order to form a basis for comparison with the capacity derived in this paper.
In Chapter \ref{sec:Capacity} the novel derivations of the capacity for LPTV channels with ACGN are detailed and the outage capacity of slow fading NB-PLC channels is discussed. 
In Chapter \ref{sec:Simulations} numerical results are presented together with a discussion. Lastly, conclusions are provided in Chapter \ref{sec:Conclusions}.

\vspace{-0.25cm}
\chapter{Preliminaries}
\label{sec:Preliminaries}
\vspace{-0.1cm}
\vspace{-0.15cm}
\section{Notations}
\label{subsec:Pre_Notations}

In the following we denote vectors with lower-case boldface letters, e.g., $\textbf{x}$; the $i$-th element of a vector $\textbf{x}$ ($i \geq 0$) is denoted with $(\textbf{x})_i$. Matrices are denoted with upper-case boldface letters, e.g., $\textbf{X}$; the  element at the $i$-th row and the $j$-th column ($i,j \geq 0$) of a matrix $\textbf{X}$ is denoted with $(\textbf{X})_{i,j}$.
$(\cdot)^H$ and $(\cdot)^T$ denote the Hermitian transpose and the transpose, respectively.
We use ${\rm {Tr}}\left(\cdot\right)$ to denote the trace operator, $\left|\cdot\right|$ to denote the absolute value when applied to scalars, and the determinant operator when applied to matrices, $\left\lfloor a \right\rfloor$ to denote the largest integer not greater than $a$, and $a^+$ to denote $\max(0,a)$. We also use $a\% b$ to denote the remainder of $a$ when divided by $b$, and $\star$ to denote the convolution operator.
Lastly, $\mathds{Z}$, $\mathds{N}$, and $\mathds{R}$ denote the set of integers, the set of non-negative integers, and the set of real numbers, respectively, $\delta[\cdot] $ denotes the Kronecker delta function,  ${\bf I}_N$ denotes the $N \times N$ identity matrix, and $\E\big\{ \cdot \}$ denotes the stochastic expectation.

\vspace{-0.15cm}
\section{Cyclostationary Stochastic Processes and the Decimated Components Decomposition}
\label{subsec:Pre_Cyclostationarity}

A real-valued DT process $x[n]$ is said to be \textit{wide-sense second order cyclostationary} (referred to henceforth as cyclostationary) if both its mean value and autocorrelation function are periodic with some period, $N_0$, i.e.,
$\E\big\{x[n]\} = \E\big\{x[n + N_0]\}$,
and
$c_{xx}(n,l) \triangleq \E\big\{x[n + l]x[n]\} = c_{xx}(n + N_0,l)$, see also  \cite[Ch. 1]{Gardner:94}, \cite[Sec. 3.2.2]{Gardner:06}.

The time-domain DCD \cite[Sec. 3.10.2]{Gardner:06}, \cite[Sec. 17.2]{Giannakis:98} transforms a scalar cyclostationary process $x[m]$ with period $N_0$ into a multivariate stationary process of size $N_0$, represented as ${\bf{x}}[n] = \big[ {x_0}[n],{x_1}[n],\ldots,{x_{N_0 - 1}}[n] \big]^T$, where ${x_q}[n] = x[nN_0 + q]$.
The process $x[m]$ can be obtained from the vector process ${\bf{x}}[n]$ via
\begin{equation}
\label{eqn:Decomp_TD0}
x[m] = \sum\limits_{i = 0}^{N_0 - 1} {\sum\limits_{l= -\infty}^{\infty} {{x_i}[l]\delta [m - i - lN_0]} }.
\end{equation}

Wide-sense second order stationary of the multivariate process ${\bf{x}}[n]$ follows since the cyclostationarity of $x[n]$ implies that
$\E\left\{ {{x_i}[n]} \right\} = \E\left\{ {x[i + nN_0]} \right\} = \E\left\{ {x[i]} \right\}$,
and
${c_{{x_i}{x_j}}}(n,l) = \E\left\{ {{x_i}[n + l]x_j[n]} \right\} = {c_{xx}}(j,i - j + lN_0)$,
which depends on $l$ but not on $n$.
We note that the DCD is the implementation in DT of the translation series representation (TSR) originally introduced in \cite{Gardner:75} for CT cyclostationary signals.
Another common transformation of scalar cyclostationary signals into an equivalent multivariate stationary process is the  harmonic series representation (HSR) \cite{Gardner:75}, sometimes referred to as the sub-band components decomposition (SBD) \cite[Sec. 17.2]{Giannakis:98}.
The SBD transforms $x[m]$ into a multivariate stationary process of size $N_0$, represented as $\widetilde{\bf{x}}[n] = \big[ {{{\widetilde x}_0}[n],{{\widetilde x}_1}[n],\ldots,{{\widetilde x}_{N_0 - 1}}[n]} \big]^T$, where with $\alpha _k \triangleq \frac{k}{N_0}$, we have ${\widetilde x_k}[n] = \left( {x[n]{e^{j{2\pi \alpha _k}n}}} \right) \star {h_o}[n]$, with ${h_o}[n] = \frac{1}{N_0}{\rm{sinc}}\left( {\frac{n}{N_0}} \right)$ being an ideal low-pass filter (LPF) whose frequency response satisfies $H_o(\omega)=1$ for $\left| \omega  \right| < \frac{\pi }{N_0}$, and zero otherwise.
The scalar process $x[m]$ can be obtained from the vector process ${\widetilde{\bf{x}}}[n]$ via $x[m] = \sum\limits_{k = 0}^{N_0 - 1} {{{\widetilde x}_k}[m]{e^{ - j{2\pi \alpha _k}m}}} $.

\vspace{-0.25cm}
\chapter{Channel Model and Related Works} 
\label{sec:Model}
\section{The Noise Model for NB-PLC}
\label{subsec:Pre_Noise}

Due to the relatively long symbol duration in NB-PLC transmissions, the periodic properties of the noise cannot be ignored
~\cite{Evans:12, Katayama:06}.
Two important works, \cite{Katayama:06} and \cite{Nassar:12}, proposed relevant passband ACGN models for NB-PLC noise which account for the periodic time domain features. These two models, referred to herein as the ``Katayama model" and the ``Nassar model", are also referenced in NB-PLC standards, see e.g., \cite{IEEE:13}, and are the baseline models for the numerical evaluations. We now briefly review these models:

{\em The Katayama noise model} proposed in \cite{Katayama:06}  models the NB-PLC noise as a real, passband, colored cyclostationary Gaussian process, $w[n]$, with $\E\{w[n]\}=0$, and with a periodic time-varying autocorrelation function $c_{ww}(n,l)$. For modeling $c_{ww}(n,l)$, the work \cite[Sec. III-B]{Katayama:06} defines $L_{noise}$ as the number of noise classes, each class $i$, $i = 0,1,\ldots, L_{noise}-1$, is represented by a sine function parametrized by three parameters: the magnitude, ${A_i}$, the power to which the sine component is raised, $\kappa_i$, and the phase, $\Theta _i$.
%
Letting $T_{AC}$ denote the cycle duration of the mains voltage, the DT noise is obtained by sampling the CT noise at instances spaced $T_{samp}$ apart.
The cyclic period of $c_{ww}(n,l)$ is thus $N_{noise}= \frac{T_{AC}}{T_{samp}}$, where
for simplicity we assume that $T_{samp}$ is such that $N_{noise}$ can be approximated as an integer.
Lastly, letting $\alpha_1$ be the coefficient of decay of $c_{ww}(n,l)$ vs. $l$ at any given time $n$,
$c_{ww}(n,l)$ is given by \cite[Eq. (17)]{Katayama:06}:
\begin{equation}
{c_{ww}}(n,l) = \frac{{{{\sum\limits_{i = 0}^{{L_{noise}} - 1} {{A_i}\left| \sin \left( {\pi \frac{n}{{{N_{noise}}}} + {\Theta _i}} \right) \right|} }^{{\kappa_i}}}}}{{1 + {{\left( {\frac{{2\pi l{T_{samp}}}}{\alpha_1 }} \right)}^2}}}. \label{eqn:NBNoise_Cor1}
\end{equation}
Observe that the correlation between the noise samples is inversely proportioned to the square of the lag value $l$.

{\em The Nassar noise model} proposed in \cite{Nassar:12}  models the NB-PLC noise as the output of an LPTV filter with a white Gaussian stochastic process (WGSP) input. To that aim, the work in \cite{Nassar:12} simultaneously filters the input WGSP by a set of $M$ LTI spectral shaping filters $\left\{h_i[l]\right\}_{i=1}^{M}$.
The time axis is divided into blocks of length $N_{noise}$, which denotes the period of the noise statistics, and each block is divided into $M$ intervals.
At each time interval the noise signal is taken from the output of one of the $M$ filters, with filter selection changing periodically.
Define $0=n_0 < n_1 < n_2 < \ldots < n_{M-1} < n_M=N_{noise}$. The $M$ interval sets $\left\{\mathcal{R}_i\right\}_{i=1}^{M}$ are defined as: $\mathcal{R}_i = \{n \in \mathds{Z}:n_{i-1} \leq (n \% N_{noise}) < n_i\}$.
Let $I[n]$ be a function which maps $n$ into the index of the interval into which $n$ belongs, i.e., if $n \in \mathcal{R}_{i_0}$ then $I[n] = i_0$, and let $\upsilon [n]$ be a zero mean, unit variance WGSP.
Define $\varepsilon _i [n] \triangleq \sum\limits_{l =  - \infty }^\infty  {{h_i}[n - l]\upsilon [l]}$, $i \in \{1,2,\ldots,M\}$. It follows that the ACGN $w[n]$ is given by $w[n] = \varepsilon _{I[n]} [n]$. The autocorrelation of $w[n]$ is obtained by
%
${c_{ww}}\left( {n,l} \right)
= \sum\limits_{{m_1} =  - \infty }^\infty  {h_{I[n+l]}}\left[ {{m_1}} \right] \cdot {h_{I[n]}}\left[ {{m_1} - l} \right]$.
%
If the filters $h_i[l]$, $i=1,2,\ldots,M$, are finite impulse response (FIR) filters, it follows that the temporal correlation is finite, i.e., $\exists L_{corr} >0$ such that $c_{ww}(n,l) = 0$, $\forall \left|l\right| \ge L_{corr}$.

\vspace{-0.25cm}
\section{The CTF for NB-PLC Channels}
\label{subsec:Pre_Channel}

Following \cite{Dostert:11, Evans:12, Katayama:08} the NB-PLC channel is modeled as a linear transformation of the input signal.
The impulse response of the power line channel is obtained by characterizing the impedances of the different electrical devices plugged into the network \cite{Dostert:11}.
Typically, these impedances depend on the level of the electric power, and are therefore periodic with a period of either half the mains period, or equal to the mains period \cite{Dostert:11, Evans:12, Katayama:08}.
To accommodate these periodic variations, the NB-PLC CTF is modeled as an LPTV system whose period is equal to the mains period \cite{Dostert:11, Evans:12, Katayama:08}.
We note that the LPTV model for the PLC CTF was originally proposed in \cite{Corripio:06} for the frequency range 1-20 MHz, and was later confirmed to hold also for the NB-PLC frequency range in \cite{Dostert:11} and \cite{Katayama:08}.
%
We use $N_{ch}$  to denote the period of the DT CTF, and $L_{isi}$ to denote the maximal memory of the channel, over all time instances $n$ in one period of $N_{ch}$ consecutive samples ($L_{isi}$ is assumed finite). The CTF coefficients at any time instance $n$ are given by $\left\{g[n,l]\right\}_{l=0}^{L_{isi}-1}$, and satisfy $g[n,l] = g[n+N_{ch},l]$.

Lastly, we note that the CTF varies also due to appliances being plugged into or out of the network, or switched on/off, and  due to topology changes of the physical power line network.
These variations are non-periodic long-term variations, and are not reflected in the periodic CTF model used in the present work.
We will briefly discuss the performance implications of the long-term variations in Section \ref{subsec:Cap_Fading}.

\vspace{-0.25cm}
\section{The Overall NB-PLC Channel Model}
\label{subsec:Pre_Overall}
Since in NB-PLC, the period of the noise statistics and the period of the CTF are harmonically related \cite{Katayama:08}, then, with a proper selection of the sampling rate, these periods are also harmonically related in the resulting DT NB-PLC channels\footnote{As described in \cite{Evans:12}, modern NB-PLC systems are commonly synchronized with the zero-crossing of the AC cycle. In fact, sampling at a rate which is an integer multiple of half the AC cycle, results in $N_{ch} = 2N_{noise}$ \cite{Dostert:11, Corripio:06}. In this manuscript we shall allow a general  relationship between the periods of the CTF and of the noise statistics.}.
Denote the least common multiple of $N_{noise}$ and $N_{ch}$ with $N_{lcm}$.
We note that in this work, the periods of the noise and of the CTF need not be harmonically related, but the sampling period is assumed to be harmonically related with $N_{lcm}$.
 Letting $x[n]$ denote the transmitted signal and $w[n]$ denote the additive noise, the signal received over the DT NB-PLC channel is modeled as ($n \in \mathds{Z}$)
\begin{equation}
\label{eqn:ChannelModel1}
r[n] = \sum\limits_{l = 0}^{{L_{isi}} - 1} {g[n,l]x[n - l]}  + w[n].
\end{equation}
Note that as we use the bandpass signal model, then all the signals are real.
We  assume that the temporal correlation of $w[n]$ is finite\footnote{This condition is satisfied by the ACGN noise model  proposed in \cite{Nassar:12}, when FIR filters are used. For the noise model in \cite{Katayama:06} we observe from \eqref{eqn:NBNoise_Cor1} that while $c_{ww}(n,l) \neq 0$ for all $l$, for large enough $\left|l\right|$ we have $c_{ww}(n,l) \approx 0$ and we can define an arbitrarily large value for which the correlation becomes small enough such that it can be neglected. The capacity of NB-PLC with the exact Katayama noise model is discussed in Section \ref{subsec:Cap_LPTV_Kata}.}, i.e., $\exists L_{corr} >0$ such that $c_{ww}(n,l) = 0$, $\forall \left|l\right| \ge L_{corr}$.
We also assume that the noise samples are not linearly dependent, i.e., no noise sample can be expressed as a linear combination of other noise samples.

\vspace{-0.25cm}
\section{A Practical Transmission Scheme for NB-PLC Based on \cite{Katayama:09} and \cite{Lampe:13}}
\label{subsec:Cap_Related}

The periodicity and frequency selectivity of narrowband power line channels can be handled by partitioning the time-frequency space into cells such that within each cell, the channel characteristics are approximately constant. This time-frequency partitioning is the underlying principle of the scheme proposed in \cite{Katayama:09} for NB-PLC channels modeled as an LTI filter with ACGN, and of the scheme proposed in \cite{Lampe:13} for broadband PLC channels, modeled as an LPTV filter with AWGN (\cite{Lampe:13} included a footnote which explained how the scheme can be adapted for ACGN). In the following we detail the extension of this design principle to LPTV channels with ACGN:
The time-frequency space is partitioned into cells.
The temporal duration of each cell, denoted $N_{sym}$, is set to an integer divisor of $N_{lcm}$, i.e., letting $N_p$ be a positive integer we write $N_{sym} = \frac{N_{lcm}}{N_p}$.
Let $B_{tot}$ be the overall bandwidth available for transmission and $N_{sc}$ be the number of subcarriers.
The frequency width of each cell, $B_{cell}$, is therefore $\frac{B_{tot}}{N_{sc}}$.
These assignments divide the $N_{lcm} \times B_{tot}$ time-frequency space into $N_p \times N_{sc}$ cells.
If $N_{sym}$ and $B_{cell}$ are set to be smaller than the coherence length and the coherence bandwidth of the channel\footnote{As in \cite{Corripio:06}, we use the terms coherence length and the coherence bandwidth to denote the range in time and frequency, respectively, in which the channel characteristics can be considered to be approximately static.}, respectively, then during the transmission of a symbol (of length $N_{sym}$) the channel at each cell may be treated as static and frequency non-selective, which facilitates the application of OFDM signal design as applied to LTI channels with stationary noise.
We henceforth refer to this scheme as {\em time-frequency (TF) OFDM}.
%
Note that as this scheme {\em inherently limits} the length of the OFDM symbol, $N_{sym}$, to be shorter than the coherence length of the channel, the cyclic prefix (CP) inserted between subsequent OFDM symbols causes a non-negligible reduction in the spectral efficiency\footnote{As a numerical example we note that the length of the  CP used in the IEEE P1901.2 standard \cite{IEEE:13} is 55 $\mu s$, while the shortest coherence time according to, e.g, \cite[Sec. IV-A]{Corripio:06} is 600 $\mu s$. Thus, the rate loss due to restricting the symbol duration to be less than the coherence time can be as high as $9 \%$.}.

At each time interval of duration $N_{sym}$, the scheme of \cite{Katayama:09} transmits a single OFDM symbol corresponding to $N_{sc}$ subcarriers appended with  $N_{cp}$ CP samples, where $N_{cp}$ is larger than the length of the ISI.
In order to accommodate the CP samples, $N_{sym}$ must satisfy $N_{sym} \geq 2N_{sc} + N_{cp}$.
To evaluate the achievable rate of the TF-OFDM scheme, let $\left|G[m,k]\right|$ and $\sigma_{m,k}^2$ denote the average channel gain (i.e., the average magnitude of the DFT of the CTF) and the average noise energy, respectively, at the $k$-th frequency cell of the $m$-th time cell, $k\in \{0,1,\ldots,N_{sc}-1\}$, $m \in \{0,1,\ldots,N_p-1\}$.
Define $\gamma_{m,k} \triangleq \frac{\left|G[m,k]\right|^2}{\sigma_{m,k}^2}$ and $\mathcal{Q} \triangleq \left\{\gamma_{m,k}\right\}_{m=0,k=0}^{N_p-1, N_{sc} -1}$, and let $x[n]$ denote the transmitted signal.
For a given average power constraint $\rho$,  $\frac{1}{N_{lcm}}\sum\limits_{n=0}^{N_{lcm}-1}\E\left\{|x[n]|^2 \right\}\le \rho$, let $\Delta_O$ be the solution to the equation
$\frac{1}{N_p N_{sc}}\sum\limits_{m=0}^{N_p-1}{\sum\limits_{k=0}^{N_{sc}-1}\left(\Delta_O - \gamma_{m,k}^{-1}\right)^+}= \rho$.
From the derivation of the achievable rate for OFDM signals with finite block length in \cite{Carmon:13},  it follows that the achievable rate of the TF-OFDM scheme in bits per channel use is given by
\begin{equation}
\label{eqn:CapOFDM4}
R_{TF-OFDM}\left(\mathcal{Q}\right) = \frac{1}{N_{lcm}} \sum\limits_{m=0}^{N_p-1}{ \sum\limits_{k=0}^{N_{sc}-1}\left(  \log\left(\Delta_O\cdot \gamma_{m,k}\right)\right)^+}.
\end{equation}
%

\vspace{-0.25cm}
\chapter{The Capacity of DT NB-PLC Channels} 
\label{sec:Capacity}
In the following we derive the capacity of DT NB-PLC channels by transforming the original scalar model of \eqref{eqn:ChannelModel1} into a MIMO model using the DCD.
%
Define $g_m[l] \triangleq g[m,l]$, 
 $L \triangleq \max\left\{L_{corr}, L_{isi}\right\}$, $K_{min} \triangleq \left\lceil \frac{L}{N_{lcm}}\right\rceil$, and let $K$ be an arbitrary positive integer s.t. $K > K_{min}$.
Let $N^{(K)} = K N_{lcm}$ denote an interval whose length is equal to $K$ common periods of duration $N_{lcm}$.
In the transformed MIMO model $N^{(K)}$ represents the dimensions of the input vector signal, and $M^{(K)} \triangleq N^{(K)} - L + 1$ represents the dimensions of the output vector signal.
Lastly, define the $M^{(K)}\times1$ vector ${\bf w}[n]$, s.t. $\left({\bf w}[n]\right)_i=w[nN^{(K)}+i+L-1]$, $i \in \{0,1,\ldots,M^{(K)}-1\} \triangleq \mathcal{M}^{(K)}$, and the $M^{(K)} \times M^{(K)}$ matrix ${\bf C}^{(K)}_{\bf ww}[n] \triangleq \E\{{\bf w}[n]\left({\bf w}[n]\right)^T\}$.
%
%
\begin{lemma}
\label{pro:NoiseCorr}
 ${\bf{C}}^{(K)}_{\bf {ww}}[n]$ is independent of $n${\normalfont :} ${\bf{C}}^{(K)}_{\bf {ww}}[n] = {\bf{C}}_{\bf {ww}}^{(K)}$.
\end{lemma}
{\em Proof:} At each $n$, ${\bf{C}}^{(K)}_{\bf {ww}}[n]$ is obtained as
\begin{align}
 \left( {{\bf{C}}^{(K)}_{\bf ww}}[n] \right)_{u,v}
	&= \E\big\{ w\left[n N^{(K)} + u + L - 1\right]
	w\left[n N^{(K)} + v + L - 1\right] \big\} \notag \\
	&= {c_{ww}}\left( {v+L-1,u - v} \right) \triangleq {\left({{\bf{C}}^{(K)}_{\bf ww}} \right)_{u,v}}, \label{eqn:Cap_NoiseCorrMat1}
\end{align}
$ u,v \in \mathcal{M}^{(K)}$. 
$\blacksquare$
\smallskip

Note that ${{\bf{C}}^{(K)}_{\bf ww}}$ is full rank as the samples of $w[m]$ are not linearly dependent.
Define the $M^{(K)}\times N^{(K)}$ matrix ${\bf{G}}^{(K)}$ such that $\forall u \in  \mathcal{M}^{(K)}$ and $\forall v \in \left\{0,1,\ldots,N^{(K)}-1\right\}\triangleq \mathcal{N}^{(K)}$,
$\big({\bf{G}}^{(K)}\big)_{u,v}=g_{u+L-1}[L-1-v+u]$ if $0 \leq v-u < L$ and $\big({\bf{G}}^{(K)}\big)_{u,v}=0$ otherwise,
i.e.,
\begin{equation}
\label{eqn:ChannelMat1}
  {\bf{G}}^{(K)} \!\!\triangleq \!
  \left[ {\begin{array}{*{20}{c}}
{g_{{L} - 1}[{L} - 1]}& \cdots &{g_{{L} - 1}[0]}& \cdots &0\\
 \vdots & \ddots &{}& \ddots & \vdots\\
0& \cdots &{g_{N^{(K)}- 1} }[{L} - 1]& \cdots &{g_{N^{(K)}- 1} }[0]
\end{array}} \right],
\end{equation}
where for $l \ge L_{isi}$ we set $g_m[l]=0$.
Define ${\bf G}^{(K)}_w \triangleq \left({\bf{C}}^{(K)}_{\bf {ww}}\right)^{-\frac{1}{2}}{\bf G}^{(K)}$,
and let ${\lambda}^{(K)}_k$, $k \in \mathcal{N}^{(K)}$, denote the $k$-th eigenvalue of ${\bf{\Gamma}}^{(K)} \triangleq \left({\bf{G}}^{(K)}_w\right)^T{\bf{G}}^{(K)}_w$.
%
%
Our main result is summarized in the following theorem:
\begin{theorem}
\label{thm:CapacityLPTV}
Consider an information signal $x[n]$, subject to a time-averaged per-symbol power constraint
\begin{equation}
\frac{1}{N}\sum\limits_{n=0}^{N-1}\E\{|x[n]|^2\} \leq \rho, \label{eqn:Cap_Constraint}
\end{equation}
received over the LPTV channel with ACGN \eqref{eqn:ChannelModel1}. Define the positive integer $K$ s.t.  $K > K_{min}$.
Select $\Delta^{(K)}$ such that $\sum\limits_{k=0}^{N^{(K)}-1}\Big(\Delta^{(K)} - \left({\lambda}^{(K)}_k\right)^{-1}\Big)^+=N^{(K)} \cdot \rho$, and define
 \begin{equation}
\label{eqn:AR_LPTV_Them}
{R}_K \triangleq \frac{1}{2N^{(K)}}\sum\limits_{k=0}^{N^{(K)}-1}{\left(\log _2\left(\Delta^{(K)} {\lambda}^{(K)}_k \right) \right)^+}.
\end{equation}
The capacity of the channel \eqref{eqn:ChannelModel1} with power constraint \eqref{eqn:Cap_Constraint} is given by the limit
\begin{equation}
\label{eqn:Cap_LPTV_Thm}
C_{LPTV-ACGN} = \mathop {\lim }\limits_{K \to \infty }{R}_K.
\end{equation}				
\end{theorem}
{\em Proof:} See Appendix \ref{app:Proof1}.

The capacity expression in \eqref{eqn:Cap_LPTV_Thm} is characterized via a limit. In the following we present an alternative capacity expression which avoids the use of a limit.
Define $N_0 \triangleq N^{(K_{min})}$, and let $\tilde{\bf w}[n]$ be an $N_0 \times 1$ vector whose elements are given by $\left(\tilde{\bf w}[n]\right)_i = w\left[nN_0 + i\right]$, $i \in \{0,1,\ldots,N_0-1\} \triangleq {\mathcal{N}}_0$.
Next, define the $N_0 \times N_0$ matrices ${\bf H}[l]$, $l \in \{0,1\}$, such that $\forall u,v \in \mathcal{N}_0$, ${\left( {\bf H}[0] \right)_{u,v}}= {g_u[u - v]}$ for ${0 \le u - v < L}$ and ${\left( {\bf H}[0] \right)_{u,v}}= 0$ otherwise, and $\left({\bf H}[1] \right)_{u,v}= g_{u}[N_0 + u - v]$ for $1 - N_0 \le u - v < L - N_0$ and $\left( {\bf H}[1] \right)_{u,v}=0$ otherwise, i.e.,
\begin{align*}
{\bf H}[0] &\triangleq \left[ {\begin{array}{*{20}{c}}
{g_0[0]}&\cdots &0&  \cdots &0\\
 \vdots & \ddots &{}&\ddots& \vdots \\
{g_{{L} - 1}[{L} - 1]}& \cdots &{g_{{L} - 1}[0]}& \cdots &0\\
 \vdots & \ddots &{}& \ddots & \vdots\\
0& \cdots &{g_{N_0 - 1}}[{L} - 1]& \cdots &{g_{N_0 - 1}}[0]
\end{array}} \right],\\
{\bf H}[1] &\triangleq \left[ {\begin{array}{*{20}{c}}
0& \cdots &0&{g_0[{L} - 1]}& \cdots &{g_0[1]}\\
 \vdots &{}& \vdots &{}& \ddots & \vdots \\
0& \cdots &0&0&{}&{g_{{L} - 2}[{L} - 1]}\\
0& \cdots &0&0& \cdots &0\\
 \vdots &{}& \vdots & \vdots &{}& \vdots \\
0& \cdots &0&0& \cdots &0
\end{array}} \right].
\end{align*}
Let ${\bf C}_{\tilde{\bf w}\tilde{\bf w}} (n,l) \triangleq \E\left\{\tilde{\bf w}[n+l]\left(\tilde{\bf w}[n]\right)^T\right\} = {\bf C}_{\tilde{\bf w}\tilde{\bf w}} (l)$, where the last equality follows as $\tilde{\bf w}[n]$ is a multivariate stationary process, since it is obtained from $w[m]$ using the DCD. Moreover, it follows from the finite correlation of $w[m]$ that $\left({\bf C}_{\tilde{\bf w}\tilde{\bf w}} (l)\right)_{u,v} = 0$, $\forall |l| >1$, $\forall u,v \in \mathcal{N}_0$.
For all $\omega \in \left[-\pi,\pi\right)$, define the $N_0 \times N_0$ matrices $\tilde{\bf H}(\omega)$, ${\bf S}_{\tilde{\bf w} \tilde{\bf w}}(\omega)$, and ${\bf \Sigma}(\omega)$, s.t.  $\left(\tilde{\bf H}(\omega)\right)_{u,v} \triangleq \sum\limits_{l=0}^1\left({\bf H}[l]\right)_{u,v}e^{-j\omega l}$, $\left({\bf S}_{\tilde{\bf w} \tilde{\bf w}}(\omega)\right)_{u,v}\triangleq \sum\limits_{l=-1}^1 \left({\bf C}_{\tilde{\bf w}\tilde{\bf w}} (l)\right)_{u,v}e^{-j\omega l}$, and ${\bf \Sigma}(\omega) \triangleq \left(\tilde{\bf H}(\omega)\right)^{H}{\bf S}_{\tilde{\bf w} \tilde{\bf w}}^{-1}(\omega)\tilde{\bf H}(\omega)$.
Lastly, let $\tilde{\lambda} _0 (\omega),\tilde{\lambda} _1 (\omega), \ldots , \tilde{\lambda} _{N_0-1} (\omega)$ denote the eigenvalues of ${\bf \Sigma}(\omega)$, $\omega \in \left[-\pi,\pi\right)$.
%
%
\begin{theorem}
\label{thm:CapacityLPTV2}
Let $\tilde{\Delta}$ be the unique solution to the equation
$ \frac{1}{2\pi }\sum\limits_{k=0}^{N_0 -1} \int\limits_{\omega = -\pi}^{\pi} \left(\tilde{\Delta} - \left(\tilde{\lambda} _k (\omega)\right)^{-1}\right)^+ d\omega =  \rho\cdot N_0$.
The capacity of the LPTV channel with ACGN \eqref{eqn:ChannelModel1} with power constraint \eqref{eqn:Cap_Constraint} is given by
\begin{equation}
C_{LPTV-ACGN} = \frac{1}{4\pi N_0}\sum\limits_{k=0}^{N_0 -1} \int\limits_{\omega = -\pi}^{\pi} \left(\log \left({\tilde{\Delta}} \cdot {\tilde{\lambda} _k (\omega)}\right)\right)^+ d\omega. \label{eqn:Cap_LPTV_Thm2}
\end{equation}				
Furthermore, the capacity-achieving input signal is a Gaussian zero-mean cyclostationary process.
\end{theorem}
{\em Proof:} See Appendix \ref{app:Proof2}.

{\bf Comment 1.}
Thm. \ref{thm:CapacityLPTV} and Thm. \ref{thm:CapacityLPTV2} state the capacity of DT NB-PLC channels modeled as an LPTV system with ACGN.
Note that LTI channels with ACGN are a special case of LPTV channels with ACGN obtained by letting $N_{ch} = 1$ and $N_{lcm} = N_{noise}$. Therefore, Thm. \ref{thm:CapacityLPTV} and Thm. \ref{thm:CapacityLPTV2} also apply to DT LTI channels with ACGN.

In the following subsections we discuss the consequences of our results in Thm. \ref{thm:CapacityLPTV} and Thm. \ref{thm:CapacityLPTV2}.

\vspace{-0.15cm}
\section{Capacity of Narrowband Powerline Channels with the Katayama Noise Model}
\label{subsec:Cap_LPTV_Kata}

Thm. \ref{thm:CapacityLPTV} relies on the assumption that the temporal correlation is finite. In the following we show that this assumption can be relaxed to the asymptotic independence assumption $\mathop {\lim }\limits_{|l| \to \infty } {c_{ww}}\left( {n,l} \right) = 0$, as is the case in the Katayama noise model \cite{Katayama:06}. This result is stated in the following corollary:
\begin{corollary}
\label{cor:CapacityLPTV_KATA}
The capacity of NB-PLC channels with Katayama noise is obtained by
\begin{equation}
\label{eqn:Cap_Kata}
C_{KATA}=\mathop {\lim }\limits_{L \to \infty }R_{2^L}.
\end{equation}
\end{corollary}
{\em Proof:}
Let $C_{KATA}$ denote the capacity of the NB-PLC channel subject to the Katayama noise model. Let $K=2^L$ and consider $L_{corr}$ large s.t. $L=\max\left\{L_{corr}, L_{isi}\right\} = L_{corr}$. Let $w_L[n]$ be an ACGN whose autocorrelation function is $c_{w_L w_L}(n,l) = c_{ww}(n,l)$ for all $n \in \mathds{Z}$ and $|l| < L$, and zero otherwise. As $L$ increases, the process $w_L[n]$ approaches $w[n]$ in distribution and thus $C_{LPTV-ACGN} \rightarrow C_{KATA}$. Thus, $\mathop {\lim }\limits_{L \to \infty }R_{2^L} = C_{KATA}$.
%
$\blacksquare$

\vspace{-0.15cm}
\section{Guidelines for a Practical Capacity-Achieving Transmission Scheme with Full CSI}
\label{sec:TxScheme}
We now discuss the practical implementation of the capacity-achieving transmission scheme used in the proof of Thm. \ref{thm:CapacityLPTV} (see Appendix \ref{app:Proof1}). Let $d[t] \in \{0,1\}$ denote an i.i.d. data stream with equal probabilities, and let $\rho$ denote the average power constraint, both are provided as inputs to the transmitter.

In the proof of Thm. \ref{thm:CapacityLPTV}, detailed in Appendix \ref{app:Proof1}, we show that the LPTV scalar channel with ACGN can be transformed into a static MIMO channel without ISI and with AWGN. Therefore, when both the transmitter and the receiver know $g[m,l]$ and $c_{ww}(m,l)$, the capacity-achieving code design for static MIMO channels with AWGN \cite[Ch. 7]{Tse:05} can be used.
For a large fixed $N^{(K)}\gg L = \max\left\{L_{corr}, L_{isi}\right\}$, both the transmitter and the receiver can use the CSI to compute  ${\bf{G}}^{(K)}$ and  ${\bf C}^{(K)}_{\bf{ww}}$. The transmitter employs encoding as in the optimal transmission scheme for static MIMO channels \cite[Ch. 7.1]{Tse:05} to map the data stream $d[t]$ into a transmitted multivariate stream ${\bf x}[n]$. The physically transmitted scalar signal $x[m]$ is then obtained from ${\bf x}[n]$ using the inverse DCD.
The receiver transforms the received scalar signal $r[m]$ into the multivariate stream ${\bf r}[n]$ using the DCD and discards the first $L-1$ elements of each received vector. The data stream is recovered from ${\bf r}[n]$ using MIMO post-processing and decoding, as detailed in \cite[Ch. 7.1]{Tse:05}.
It therefore follows that the capacity-achieving scheme of Thm. \ref{thm:CapacityLPTV} is feasible in the same sense that the optimal capacity-achieving scheme for static MIMO channels with AWGN \cite[Ch. 7]{Tse:05} is feasible.

\vspace{-0.15cm}
\section{Comments on the Outage Capacity of Slow Fading LPTV Channels with ACGN}
\label{subsec:Cap_Fading}

The statistical model for the CTF of the power line channel was studied in \cite{Prassana:09}, \cite{Evans:12}, and \cite{Canete:03}. The work \cite{Canete:03} also characterized the rate of the channel variations, and showed that the CTF changes on the order of minutes.
Combining the statistical model of \cite{Prassana:09} with the long-term variations characterized in \cite{Canete:03}, the NB-PLC CTF can be modeled as a random vector, which obtains a new realization every time the network topology changes (i.e., on a scale of minutes).
This gives rise to {\em the slow fading model} for NB-PLC.
As there does not exist a generally accepted statistical model for the NB-PLC CTF, we formulate the problem as communications over a slow-fading ACGN with a general random LPTV CTF without explicitly stating a probability density function (PDF) for the CTF.
We then present a general expression for the outage capacity for such slow-fading PLC channels together with an upper bound on this capacity.

Define the set of $N_{ch}\cdot L_{isi}$ channel coefficients $\mathcal{G} \triangleq \left\{g[n,l]\right\}_{n=0,l=0}^{N_{ch}-1,L_{isi}-1}$. This set constitutes an  $N_{ch}\cdot L_{isi} \times 1$ multivariate RV with a PDF $f_{\mathcal{G}}(\cdot)$.
For a specific realization of $\mathcal{G}$, the received signal is given by \eqref{eqn:ChannelModel1}.
Since the receiver knows the channel coefficients, $\left\{g[n,l]\right\}$, and the autocorrelation function of the noise, $c_{ww}(n,l)$, it can form the equivalent MIMO channel matrix ${\bf{G}}^{(K)}$ via \eqref{eqn:ChannelMat1}, and the covariance matrix of the noise ${\bf{C}}^{(K)}_{\bf ww}$ via \eqref{eqn:Cap_NoiseCorrMat1}. The receiver can now form the whitened MIMO channel matrix through ${\bf G}^{(K)}_w = \left({\bf{C}}^{(K)}_{\bf {ww}}\right)^{-\frac{1}{2}}{\bf G}^{(K)}$. For a fixed $K$, $N^{(K)} = KN_{lcm}$, $L= \max\left\{L_{corr}, L_{isi}\right\}$,
and input covariance matrix ${\bf{C}}_{{\bf{xx}}}$, the achievable rate is obtained as
$R_{F}({\bf{C}}_{{\bf{xx}}}, \mathcal{G}, N^{(K)} ,L) =  \frac{1}{2N^{(K)}}{{\log }_2}  \big| {{\bf{I}}_{M^{(K)}}} + {\bf{G}}^{(K)}_w{\bf{C}}_{{\bf{xx}}}\left({\bf{G}}^{(K)}_w\right)^T  \big|$ (see Eq. \eqref{eqn:Cap_LPTV4} in Appendix \ref{app:Proof1}), where the subscript $F$ is used for denoting the fact that the channels considered in this discussion are fading channels, instead of the static channel considered previously.

When the channel realizations {\em are not known to the transmitter}, the transmission rate must be fixed, and an outage may occur. The outage probability for a target rate $R_T$ and a given input covariance matrix ${\bf{C}}_{{\bf{xx}}}$ is given by
\begin{equation*}
{P_{outage}}\left({\bf{C}}_{{\bf{xx}}}, R_T, N^{(K)} ,L \right) = \Pr \left\{ {R_{F}\left({\bf{C}}_{{\bf{xx}}},\mathcal{G}, N^{(K)} ,L\right) \le R_T} \right\}.
\end{equation*}
Let $\mathcal{T}_{N^{(K)}}$ be the set of all $N^{(K)} \times N^{(K)}$ non negative definite matrices ${\bf T}$ satisfying ${\rm{Tr}}\left({\bf T}\right)  \leq N^{(K)}  \cdot \rho$.
The optimal input covariance matrix is the matrix ${\bf{C}}_{{\bf{xx}}}\in \mathcal{T}_{N^{(K)}}$ which minimizes ${P_{outage}}\left({\bf{C}}_{{\bf{xx}}}, R_T, N^{(K)} ,L \right)$.
In order to avoid cluttering, in the sequel we denote the minimal outage probability with $\Psi$, i.e.,
\begin{equation}
\Psi\triangleq \mathop {\min }\limits_{{\bf{C}}_{{\bf{xx}}} \in \mathcal{T}_{N^{(K)}} } {P_{outage}}\left({\bf{C}}_{{\bf{xx}}}, R_T, N^{(K)} ,L \right). \label{eqn:Cap_Fad2}
\end{equation}
Since the optimal ${\bf{C}}_{{\bf{xx}}}$ in \eqref{eqn:Cap_Fad2} depends on the PDF of the channel coefficients $f_{\mathcal{G}}(\cdot)$, it may be difficult to compute. We therefore provide an upper bound to $\Psi$  by selecting a specific arbitrary input covariance matrix which satisfies the power constraint. Setting ${\bf{C}}_{{\bf{xx}}} = \rho {\bf I}_{N^{(K)}}$, we obtain the following upper bound:
\begin{equation}
\Psi \le \Pr \left\{ \frac{{{\log }_2}  \big| {{\bf{I}}_{M^{(K)}}} +  \rho {\bf{G}}^{(K)}_w\left({\bf{G}}^{(K)}_w\right)^T  \big|}{2N^{(K)}} \le R_T \right\}.
\label{eqn:Cap_Fad4}
\end{equation}
Note that while \eqref{eqn:Cap_Fad4} is simpler to compute than \eqref{eqn:Cap_Fad2}, it may not be tight, depending on the PDF of $\mathcal{G}$.
\vspace{-0.25cm}
\chapter{Numerical Evaluations}
\label{sec:Simulations}

In this chapter we numerically compare the {\em capacity} we derived in Section \ref{sec:Capacity} with the {\em achievable rates} of the ad-hoc TF-OFDM scheme based on \cite{Katayama:09} and  \cite{Lampe:13}, described in Section \ref{subsec:Cap_Related}, and demonstrate the benefits of the capacity-approaching coding scheme for practical NB-PLC scenarios over the TF-OFDM scheme. The achievable rates of the TF-OFDM scheme were evaluated using \eqref{eqn:CapOFDM4}.
The discussion in Section \ref{sec:Intro} implies that there is no basis for comparison between the present work and the work of \cite{Han:12}, hence the results of \cite{Han:12} are not evaluated in this chapter.
%
%
%
%

In the numerical evaluations we consider the frequency band  up to $150$ kHz, which is in accordance with the European CENELEC regulations~\cite{CA:91}. The sampling frequency is $300$ kHz.
Two types of noise are simulated -

{\em 1)} {ACGN based on the Nassar model~\cite{Nassar:12} using two sets of typical parameters specified in the IEEE P1901.2 standard~\cite{IEEE:13}, referred to in the following as \textquoteleft${\mbox{IEEE1}}$\textquoteright ~and \textquoteleft${\mbox{IEEE2}}$\textquoteright: \\
	$\bullet$ {\textquoteleft${\mbox{IEEE1}}$\textquoteright ~corresponds to low voltage site 11 (LV11) in~\cite[Appendix G]{IEEE:13}.}	\\
	$\bullet$ {\textquoteleft${\mbox{IEEE2}}$\textquoteright ~corresponds to low voltage site 16 (LV16) in~\cite[Appendix G]{IEEE:13}.} \\
Each filter in the implementation of the LPTV noise model consists of $8$ taps, implying that $L_{corr} = 7$.
}

{\em 2)} {ACGN based on the Katayama model~\cite{Katayama:06} with two sets of typical parameters, referred to in the following as \textquoteleft${\mbox{KATA1}}$\textquoteright ~ and \textquoteleft${\mbox{KATA2}}$\textquoteright: \\
	$\bullet$ {The parameters for \textquoteleft${\mbox{KATA1}}$\textquoteright ~are taken from~\cite{Katayama:06} and are set to be $\{ {n_0}, {n_1}, {n_2} \} = \{ 0,1.91,1.57 \cdot {10^5}\} $, $\{ {\Theta _0}, {\Theta _1}, {\Theta _2} \} = \{ 0, - 6,- 35\} $ degrees, $\{ {A_0}, {A_1}, {A_2} \} = \{ 0.23, 1.38,$ $7.17\} $, and $\alpha_1 = 1.2\cdot 10^{-5}$.} \\
	$\bullet$ {The parameters for \textquoteleft${\mbox{KATA2}}$\textquoteright ~are taken from~\cite[residence 1]{Katayama:00}, and are set to be $\{ {n_0}, {n_1}, {n_2} \} = \{ 0,9.3,5.3 \cdot {10^3}\} $, $\{ {\Theta _0}, {\Theta _1}, {\Theta _2} \} = \{ 0, 128, 161\} $ degrees, $\{ {A_0}, {A_1}, {A_2} \} = \{ 0.13, 2.8, 16\} $, and $\alpha_1 = 8.9\cdot 10^{-6}$.} \\
In the capacity evaluation for the Katayama model via Corollary \ref{cor:CapacityLPTV_KATA}, $L_{corr}$ was selected to be the smallest value for which $c_{ww}(n,l)$ is less than $10^{-3}$ of its peak value. This resulted in $L_{corr}=9$ for \textquoteleft${\mbox{KATA1}}$\textquoteright ~and $L_{corr}=11$ for \textquoteleft${\mbox{KATA2}}$\textquoteright.
}

Two CTFs are used in the evaluations: a static flat channel, denoted in the following as \textquoteleft${\mbox{Flat}}$\textquoteright, and an LPTV channel, denoted in the following as \textquoteleft${\mbox{LPTV}}$\textquoteright.
As there is no universally acceptable CTF generator for NB-PLC \cite{Hoch:11}, we shall follow the approach of \cite{Prassana:09} and generate the CTF based on transmission line theory, similar to the approach used in \cite{Corripio:11} for broadband PLC.
%
The channel generator proposed in \cite{Corripio:11} was adapted to NB-PLC by setting the intermediate impedances of the model, $Z_1$, $Z_2$, and $Z_3$, to series RLC resonators with resistance of $17 k\Omega$, $8 k\Omega$, and $26 k\Omega$, respectively, capacitance of $0.6 nF$, $1 nF$, and $3.7 nF$, respectively, and inductance of $11 mH$, $2.3 mH$, and $6 mH$, respectively.
These values correspond to frequency selectivity in the band of up to $150$ kHz.
The short-term temporal variations of the channel were realized by setting the time-varying behavior of $Z_1$ and $Z_3$ to the harmonic behavior (see \cite[Pg. 168]{Corripio:11}), with a phase of $\frac{\pi}{2}$ and $\frac{\pi}{4}$, respectively, and setting the time-varying behavior of $Z_2$ to the commuted case (see \cite[Pg. 168]{Corripio:11}), with duty cycle of $\frac{1}{8}$ of the mains period. The rest of the parameters used by the channel generator were set to the default values in \cite{Corripio:11}.
Hence, the generated LPTV CTF captures the essence of the NB-PLC CTF.
For the channel generated according to the above procedure, $L_{isi}$ was taken as the largest time index after which the corresponding real-valued impulse response does not exceed $1\%$ of the maximal tap magnitude. This resulted in $L_{isi}=8$. Then, all filter coefficients beyond $L_{isi}$ were truncated, resulting in an $8$-tap filter realization.
The results are plotted for various values of input SNR defined as ${\mbox{SNR}}_{in} \triangleq \rho \Big( \frac{1}{N_{noise}}\sum\limits_{n=0}^{N_{noise}-1}c_{ww}(n,0) \Big)^{-1}$.

We determine the parameters of the TF-OFDM scheme based on the IEEE P1901.2 standard \cite{IEEE:13} applied in the CENELEC environment \cite{CA:91}. From the specification in \cite{IEEE:13} we conclude that the ratio between the period of the noise and the duration of the OFDM symbol  is $14.3$, thus we set $N_p = 14$. $N_{sym}$ is set to $\left\lfloor \frac{N_{lcm}}{N_p}\right\rfloor$, and $N_{sc}$ is set to $\lfloor \frac{1}{2}(N_{sym} - L_{isi} + 1)\rfloor$.

\begin{figure}
\centering
\scalebox{0.5}{\includegraphics[clip=true,viewport=0.75in 0.2in 7.6in 5.6in]{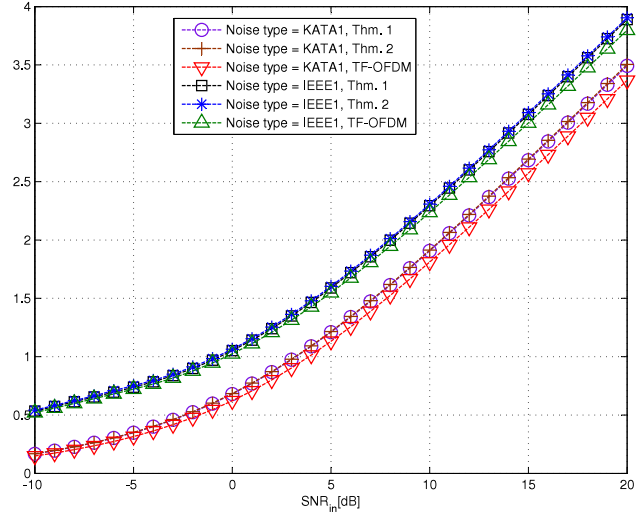}}
\vspace{-0.4cm}
  \caption{Achievable rate comparison for the \textquoteleft${\mbox{Flat}}$\textquoteright ~channel.}
	\vspace{-0.4cm}
  \label{fig:Cap_FlatKATA}
\end{figure}

{ \bf{\em{1) Evaluating the Achievable Rates for Static Flat ACGN Channels}:}}
We first evaluated the achievable rates for static flat ACGN channels ($L_{isi}=1$). The results for both noise models with the \textquoteleft${\mbox{KATA1}}$\textquoteright ~and \textquoteleft${\mbox{IEEE1}}$\textquoteright ~parameters sets are depicted in Fig. \ref{fig:Cap_FlatKATA}.
Since the channel is flat, then we can set the length of the cyclic prefix to $N_{cp}=0$, and thus $N_{sym} = 2N_{sc}$.
As expected, the numerical evaluations of the capacity derived via Thm. \ref{thm:CapacityLPTV} and that derived via Thm. \ref{thm:CapacityLPTV2} coincide.
Note that the achievable rate of the TF-OFDM scheme is slightly less than capacity since in both the \textquoteleft${\mbox{KATA1}}$\textquoteright ~scenario and the \textquoteleft${\mbox{IEEE1}}$\textquoteright ~scenario, $N_{sym}$ is larger than the coherence duration of the noise, i.e., the statistics of the noise vary within a single OFDM symbol duration.

\begin{figure}
\centering
\scalebox{0.5}{\includegraphics[clip=true,viewport=0.75in 0.2in 7.6in 5.6in]{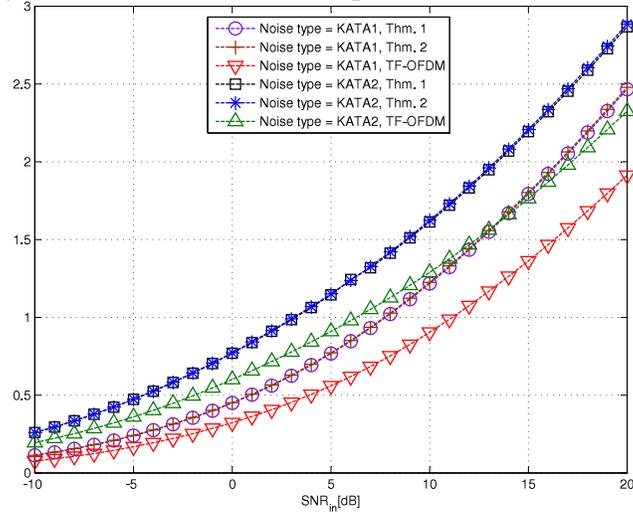}}
\vspace{-0.4cm}
  \caption{Achievable rate comparison the Katayama noise model of~\cite{Katayama:06} for the \textquoteleft${\mbox{LPTV}}$\textquoteright ~channel.}
	\vspace{-0.4cm}
  \label{fig:Cap_LPTV_KATA}
\end{figure}

\begin{figure}
\centering
\scalebox{0.5}{\includegraphics[clip=true,viewport=0.75in 0.2in 7.6in 5.6in]{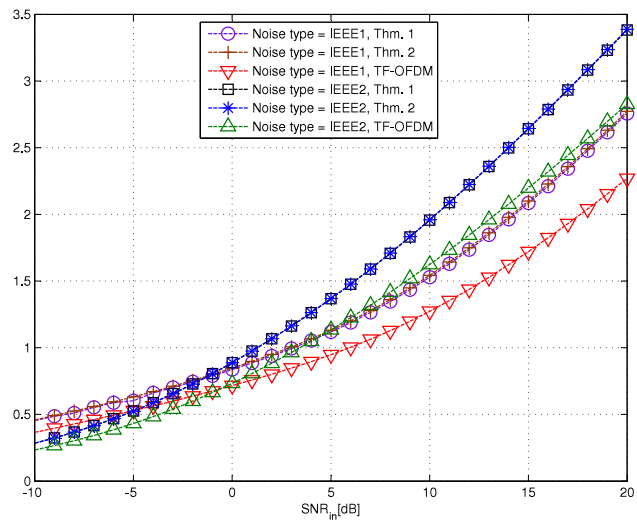}}
\vspace{-0.4cm}
  \caption{Achievable rate comparison for the LPTV noise model of~\cite{Nassar:12} for the \textquoteleft${\mbox{LPTV}}$\textquoteright ~channel.}
	\vspace{-0.2cm}
  \label{fig:Cap_LPTV_IEEE}
\end{figure}

{ \bf{\em{2) Achievable Rate Improvement for NB-PLC Channels}:}}
Next, the achievable rates were evaluated for NB-PLC channels, using the LPTV channel with ACGN model.
The results for the LPTV channel with the Katayama noise model of~\cite{Katayama:06} are depicted in Fig. \ref{fig:Cap_LPTV_KATA}, and the results for the LPTV channel with the Nassar noise model of~\cite{Nassar:12} are depicted in Fig. \ref{fig:Cap_LPTV_IEEE}.
It is clearly observed that the numerical evaluations of Thm. \ref{thm:CapacityLPTV} and Thm. \ref{thm:CapacityLPTV2} coincide, reaffirming the equivalence of the capacity expressions derived in these theorems.
As expected, the capacity stated in Thm. \ref{thm:CapacityLPTV} and in Thm. \ref{thm:CapacityLPTV2} exceeds the achievable rate of the ad-hoc TF-OFDM scheme.
Note that the rate of the TF-OFDM scheme is considerably less than the channel capacity especially at high SNRs, as at these SNRs the rate loss due to the non-negligible cyclic prefix is more dominant. 
For the \textquoteleft${\mbox{KATA1}}$\textquoteright ~noise model the loss varies from $2.9$ dB at ${\mbox{SNR}}_{in}=0$ dB to $3.5$ dB at ${\mbox{SNR}}_{in}=10$ dB, for the \textquoteleft${\mbox{KATA2}}$\textquoteright ~noise model the loss varies from $2.8$ dB at ${\mbox{SNR}}_{in}=0$ dB to $3.55$ dB at ${\mbox{SNR}}_{in}=10$ dB, for the \textquoteleft${\mbox{IEEE1}}$\textquoteright ~noise model the loss varies from $2.7$ dB at ${\mbox{SNR}}_{in}=0$ dB to $3$ dB at ${\mbox{SNR}}_{in}=10$ dB, and for the \textquoteleft${\mbox{IEEE2}}$\textquoteright ~noise model the loss varies from $2.05$ dB at ${\mbox{SNR}}_{in}=0$ dB to $2.9$ dB at ${\mbox{SNR}}_{in}=10$ dB.
We thus conclude that optimally accounting for the time-variations of the channel in the design of the transmission scheme  for NB-PLC leads to substantial SNR gains.

\chapter{Conclusions}
\label{sec:Conclusions}
\vspace{-0.1cm}
In this paper, the capacity of NB-PLC channels was derived, and the corresponding optimal transmission scheme was obtained.
The novel aspect of the work is the insight that by applying the DCD to the scalar LPTV channel with ACGN, the periodic short-term variations of the NB-PLC CTF, as well as the cyclostationarity of the noise, are converted into time-invariant properties of {\em finite} duration in the resulting MIMO channel. This was not possible with the previous approach which considered frequency-domain decomposition. 
In the numerical evaluations, a substantial rate increase is observed compared to the previously proposed practical TF-OFDM scheme obtained from \cite{Katayama:09} and \cite{Lampe:13}.
Future work will focus on adapting cyclostationary signal processing schemes to MIMO NB-PLC channels.

\begin{appendix}
\numberwithin{equation}{chapter}
\numberwithin{lemma}{chapter}
\numberwithin{claim}{chapter}
\numberwithin{definition}{chapter}

\vspace{-0.2cm}
\chapter{Proof of Thm. \ref{thm:CapacityLPTV}}
\label{app:Proof1}
\begin{figure}
	\centering
		\includegraphics[width=\textwidth]{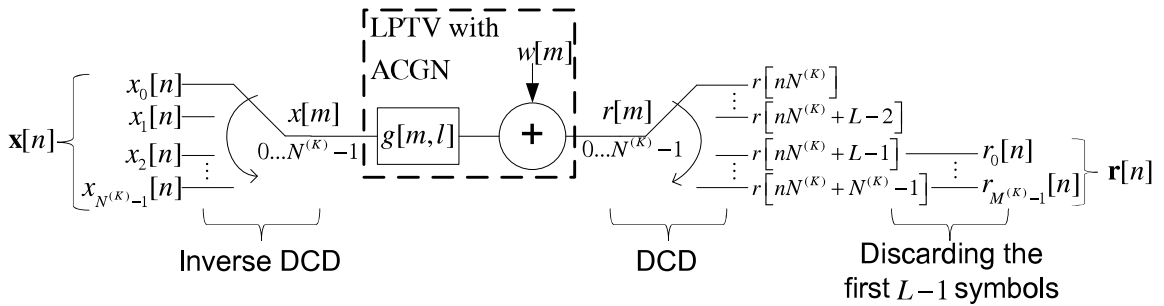}
		\vspace{-0.6cm}
	\caption{The transformed MIMO channel model obtained from the scalar DT LPTV ACGN channel by applying the inverse DCD to the input signal and the DCD to the output signal.}
	\label{fig:CAP_TDDecomp1}
	\vspace{-0.4cm}
\end{figure}

\vspace{-0.1cm}
The proof of Thm. \ref{thm:CapacityLPTV} consists of two parts: We first consider the channel \eqref{eqn:ChannelModel1} in which $L-1$ out of every $N^{(K)}$ channel outputs are discarded by the receiver, and show that the achievable rate of this channel is given in \eqref{eqn:AR_LPTV_Them}. Next, we prove that \eqref{eqn:AR_LPTV_Them} denotes the capacity of the LPTV channel with ACGN when taking $K \rightarrow \infty$.
Let us observe the decomposed polyphase channel \eqref{eqn:ChannelModel1} depicted in Fig. \ref{fig:CAP_TDDecomp1}.
Fix $K$ and define ${x_i}[n] \triangleq x\left[nN^{(K)}+i\right]$, ${r_i}[n] \triangleq r\left[nN^{(K)} + L - 1 +i\right]$, and ${w_i}[n] \triangleq w\left[nN^{(K)}+L-1+i\right]$.
For $i \in \mathcal{M}^{(K)}$ and $n$ fixed,
\eqref{eqn:ChannelModel1} can be written as
$r_i[n] = \sum\limits_{l = 0}^{{L_{isi}} - 1} g\left[n{N}^{(K)}+L-1+i,l\right]{x_{i+L-1 - l}}[n]  + {w_i}[n]$.
Since the channel coefficients are periodic with a period $N_{ch}$, which is a divisor of $N^{(K)}$, it follows that $g[n{N}^{(K)}+i+L-1,l] =  g[i+L-1,l] \equiv g_{i+L-1}[l]$. Hence, $r_i[n]$ can be written as
\begin{equation}
\label{eqn:Cap_LPTV2}
{r_i}[n] = \sum\limits_{l = 0}^{{L_{isi}} - 1} {g_{i+L-1}[l]{x_{i+L-1 - l}}[n]}  + {w_i}[n],
\end{equation}
$i \in \mathcal{M}^{(K)}$. Next, we define the $M^{(K)}\times1$ vector ${\bf r}[n]$ such that $\left({\bf r}[n]\right)_i=r_i[n]$, $i \in \mathcal{M}^{(K)}$, and the $N^{(K)}\times1$ vector ${\bf x}[n]$ such that $\left({\bf x}[n]\right)_u=x_u[n]$, $u \in \mathcal{N}^{(K)}$.
With these definitions, we can write \eqref{eqn:Cap_LPTV2} as an equivalent $M^{(K)}\times N^{(K)}$ MIMO channel
\begin{equation}
\label{eqn:Cap_LPTV3}
{\bf r}[n]={\bf G}^{(K)}\cdot{\bf x}[n] + {\bf w}[n].
\end{equation}
Note that the vectors ${\bf w}[n]$ and ${\bf w}[n+l]$, $l \neq 0$, are jointly Gaussian as all of their elements are samples of the Gaussian process $w[m]$.
Since $L \ge L_{corr}$, the vectors ${\bf w}[n]$ and ${\bf w}[n+l]$ are uncorrelated (recall that the mean is zero), and are thus statistically independent $\forall l \neq 0$.
It then follows that the multivariate noise process ${\bf w}[n]$ is i.i.d. in time and has a jointly stationary multivariate zero-mean Gaussian distribution at every time instance $n$.
Define ${{\bf{w}}_w}[n] \triangleq \left({\bf{C}}^{(K)}_{\bf {ww}}\right)^{ - \frac{1}{2}}{\bf{w}}[n]$.
Applying a noise whitening filter to the received signal, we obtain the following equivalent channel
\begin{equation}
{{\bf{r}}_w}[n] = \left({\bf{C}}^{(K)}_{\bf {ww}}\right)^{ - \frac{1}{2}}{\bf{r}}[n] = {\bf{G}}^{(K)}_{w}{\bf{x}}[n] + {{\bf{w}}_w}[n]. \label{eqn:LPTV_Cap4}
\end{equation}
Note that the noise of the equivalent channel, ${{\bf{w}}_w}[n]$, is a multivariate AWGN process with an identity correlation matrix.
With the representation \eqref{eqn:LPTV_Cap4}, the problem of communications over an LPTV channel with ACGN \eqref{eqn:ChannelModel1} is transformed into communications over a time-invariant MIMO channel with AWGN, i.i.d. in time,
with the power constraint obtained from \eqref{eqn:Cap_Constraint}. Observe that each message is transmitted via an ${N}^{(K)} \times B$ matrix ${\bf X}$, with $B$ being a large integer representing the codeword length in the MIMO channel. It thus follows from \eqref{eqn:Cap_Constraint} that ${\bf x}[n]$ must satisfy
$\frac{1}{B \cdot N^{(K)}}\sum\limits_{n=0}^{B-1}\sum\limits_{k=0}^{{N}^{(K)}-1}\E\{(x_k[n])^2\} \leq \rho$, thus $\frac{1}{B }\sum\limits_{n=0}^{B-1}\sum\limits_{k=0}^{{N}^{(K)}-1}\E\{(x_k[n])^2\} \leq N^{(K)}\cdot \rho$.
With this new MIMO average power constraint, the capacity is obtained as follows: Let $\mathcal{T}_{N^{(K)}}$ be the set of all $N^{(K)} \times N^{(K)}$ non-negative definite matrices ${\bf T}$ satisfying ${\rm{Tr}}\left({\bf T}\right)  \leq N^{(K)}  \cdot \rho$.
The capacity expression (in bits per MIMO channel use) for the equivalent channel \eqref{eqn:LPTV_Cap4} subject to the constraint $\frac{1}{B }\sum\limits_{n=0}^{B-1}\sum\limits_{k=0}^{{N}^{(K)}-1}\E\{(x_k[n])^2\} \leq N^{(K)}\cdot \rho\;$ is well known \cite[Thm. 9.1]{ElGamal:10}\footnote{We note that the theorem in \cite[Thm. 9.1]{ElGamal:10} is stated for a per-codeword constraint. However, following \cite[Ch. 7.3, pgs. 323-324]{Gallager:68} it is immediate to show that the proof of \cite[Thm. 9.1]{ElGamal:10} also holds subject to the average power constraint $\frac{1}{B \cdot N^{(K)}}\sum\limits_{n=0}^{B-1}\sum\limits_{k=0}^{{N}^{(K)}-1}\E\{(x_k[n])^2\} \leq \rho$.}:
\begin{equation}
\label{eqn:SISO_Cap5}
C_{MIMO}^{(K)} = \mathop {\max }\limits_{{\bf{C}}_{{\bf{xx}}} \in \mathcal{T}_{N^{(K)}} } \bigg\{ \frac{1}{2}{\log _2} \bigg| {{\bf{I}}_{M^{(K)}}} + {{\bf{G}}^{(K)}_w}{\bf{C}}_{{\bf{xx}}}\left({\bf{G}}^{(K)}_w\right)^T \bigg| \bigg\}.
\end{equation}

Note that in the transformation we drop $L-1$ samples out of every $N^{(K)}$ samples. As will be proven in Appendix \ref{app:Proof1a}, this loss is asymptotically negligible when $K \rightarrow \infty$, and thus, it does not affect the capacity.
As is clear from \cite[Thm. 9.1]{ElGamal:10}, the capacity of the MIMO channel \eqref{eqn:LPTV_Cap4}, stated in \eqref{eqn:SISO_Cap5}, is obtained by input vectors generated i.i.d. in time according to a zero-mean multivariate Gaussian distribution with a covariance matrix ${\bf {C}}_{\bf xx}$, which satisfies the given power constraint.
The scalar signal $x[m]$, obtained from the columns of ${\bf X}$ via the inverse DCD, satisfies $\E\big\{ x[m] \big\} =  0$ and
\begin{align*}
c_{xx}(m,l)
&= \E\Bigg\{ \bigg({\bf x}\bigg[ \left\lfloor {\frac{{m + l}}{N^{(K)}}} \right\rfloor  \bigg]\bigg) _{\left( {m + l} \right)\% N^{(K)}}\!  \Big({\bf x}\Big[ \left\lfloor {\frac{m}{N^{(K)}}} \right\rfloor  \Big]\Big) _{m\% N^{(K)}} \Bigg\} \\
&\stackrel{(a)}{=} {\left( {\bf {C}}_{\bf xx} \right)_{\left( {m + l} \right)\% N^{(K)},m\% N^{(K)}}}\cdot \delta \bigg[ \left\lfloor {\frac{{m + l}}{N^{(K)}}} \right\rfloor  - \left\lfloor {\frac{m}{N^{(K)}}} \right\rfloor  \bigg] \\
&\stackrel{(b)}{=} c_{xx}(m+N^{(K)},l),
\end{align*}
where $(a)$ follows since when $\left\lfloor {\frac{{m + l}}{N^{(K)}}} \right\rfloor  \neq \left\lfloor {\frac{m}{N^{(K)}}} \right\rfloor $, the samples are taken from random vectors (RVs) with different time indexes, and as ${\bf x}[n]$ is i.i.d. over $n$ and has a zero mean, the corresponding cross correlation is zero; and $(b)$ follows since $m$ can be replaced by $m + N^{(K)}$ without affecting the expression in step $(a)$.
This shows that the capacity-achieving input scalar signal $x[m]$ is a cyclostationary Gaussian stochastic process with period $N^{(K)}$,
%
and therefore, the average power constraint \eqref{eqn:Cap_Constraint} can be written as
\begin{equation}
\label{eqn:Constraint2}
\frac{1}{{N}^{(K)}}\sum\limits_{k=0}^{{N}^{(K)}-1}\E\{(x_k[n])^2\} \le \rho.
\end{equation}
Proceeding with the derivation, note that since each MIMO symbol vector is transmitted via $N^{(K)}$ channel uses, the achievable rate for the physical scalar channel (in bits per channel use) is given by $\frac{1}{N^{(K)}}C_{MIMO}^{(K)}$, i.e.,
\begin{equation}
\label{eqn:Cap_LPTV4}
  R_K
= \frac{1}{2N^{(K)}}\!\! \mathop {\max }\limits_{{\bf{C}}_{{\bf{xx}}} \in \mathcal{T}_{N^{(K)}} } \left\{ {\log _2} { \left| {{{\bf{I}}_{M^{(K)}}} + {{\bf{G}}^{(K)}_w}{\bf{C}}_{{\bf{xx}}}\left({\bf{G}}^{(K)}_w\right)^T}\! \right|} \right\}.
\end{equation}
It follows from \cite[Ch. 9.1]{ElGamal:10} that the eigenvectors of the matrix ${\bf{C}}_{{\bf{xx}}}^{opt}$ which maximizes \eqref{eqn:Cap_LPTV4} coincide with the eigenvectors of ${\bf{\Gamma}}^{(K)}$. We therefore write ${\bf{\Gamma}}^{(K)} = {\bf{V}}{\bf{\Lambda}}{\bf{V}}^T$, and ${\bf{C}}_{{\bf{xx}}}^{opt} = {\bf{V}}{\bf{D}}{\bf{V}}^T$, where ${\bf{\Lambda}}$ and ${\bf{D}}$ are the ${N}^{(K)} \times {N}^{(K)}$ diagonal eigenvalue  matrices for the corresponding eigenvalue decompositions (EVDs), and ${\bf{V}}$ is the ${N}^{(K)} \times {N}^{(K)}$ unitary eigenvectors matrix for the EVD of ${\bf{\Gamma}}^{(K)}$.
From \cite[Ch. 9.1]{ElGamal:10}  it also follows that the diagonal entries of ${\bf{D}}$ are obtained by ``waterfilling" on the eigenvalues of ${\bf{\Gamma}}^{(K)}$: Letting ${\lambda}_k^{(K)} \triangleq \left({\bf{\Lambda}}\right)_{k,k} \geq 0$, waterfilling leads to the assignment $\left({\bf{D}}\right)_{k,k} = \bigg(\Delta^{(K)} - \left({\lambda}_k^{(K)}\right)^{-1}\bigg)^+$, $k \in {\mathcal{N}}^{(K)}$, where $\Delta^{(K)}$ is selected s.t. ${\rm{Tr}}\left( {\bf{D}} \right) = \sum\limits_{k=0}^{{N}^{(K)}-1}\bigg(\Delta^{(K)} - \left({\lambda}_k^{(K)}\right)^{-1}\bigg)^+ = {N}^{(K)} \cdot \rho$.
Next, write the singular value decomposition (SVD) of ${{\bf{G}}^{(K)}_w}$ as ${{\bf{G}}^{(K)}_w} = {\bf{U}}{\bf{\Lambda}}_{S}{\bf{V}}^H$, where ${\bf{U}}$ is an $M^{(K)} \times M^{(K)}$ unitary matrix and ${\bf{\Lambda}}_{S}$ is an $M^{(K)} \times N^{(K)}$ diagonal matrix which satisfies ${\bf{\Lambda}}_{S}^T {\bf{\Lambda}}_{S} = {\bf{\Lambda}}$.
Note that since ${\rm {rank}}\left({\bf{G}}^{(K)}_w\right) \leq M^{(K)}$, then at most $M^{(K)}$ of the diagonal entries of ${\bf{\Lambda}}$ are non-zero.
Plugging ${\bf{C}}_{{\bf{xx}}}^{opt}$ into \eqref{eqn:Cap_LPTV4} we obtain
\begin{align}
R_K &= \frac{1}{2N^{(K)}}\log _ 2 \left|{{{\bf{I}}_{M^{(K)}}} + {{\bf{G}}^{(K)}_w}{\bf{C}}_{{\bf{xx}}}^{opt}\left({\bf{G}}^{(K)}_w\right)^T} \right|  \label{eqn:SISO_CapTemp1}\\
&\stackrel{(a)}{=}  \frac{1}{2N^{(K)}}\log _ 2 \left|{\bf{I}}_{N^{(K)}} + {\bf{D}}{\bf{\Lambda}}_{S}^T{\bf{\Lambda}}_{S} \right|  \notag\\
&\stackrel{(b)}{=}  \frac{1}{2N^{(K)}}\log _ 2 \left|{\bf{I}}_{N^{(K)}} + {\bf{D\Lambda}} \right|    \notag \\
&\stackrel{(c)}{=}\frac{1}{2N^{(K)}}\sum\limits_{k=0}^{N^{(K)}-1}{\left(\log _2\left(\Delta^{(K)} {\lambda}^{(K)} _k \right) \right)^+}, \label{eqn:SISO_Cap7}
\end{align}
where $(a)$ follows from Sylvester's determinant theorem \cite[Ch. 6.2]{Meyer:00}; $(b)$ follows from ${\bf{\Lambda}}_{S}^T {\bf{\Lambda}}_{S} = {\bf{\Lambda}}$; and $(c)$ is obtained by plugging the expressions for $\left({\bf{D}}\right)_{k,k}$ and $\left({\bf{\Lambda}}\right)_{k,k}$. Note that \eqref{eqn:SISO_Cap7} coincides with \eqref{eqn:AR_LPTV_Them} in Thm. \ref{thm:CapacityLPTV}.

Next, we prove that for $K \rightarrow \infty$, the capacity of the transformed channel \eqref{eqn:SISO_Cap7} denotes the capacity of the LPTV channel with ACGN.
Note that since the DCD leads to an equivalent signal representation for the LPTV channel with ACGN, the capacity of the equivalent model is identical to that of the original signal model.
%
In the following we show that dropping the first $L-1$ symbols out of each vector of length $N^{(K)}$ does not change the rate when $K \rightarrow \infty$.
The intuition is that since the block length is $N^{(K)} = KN_{lcm}$, where $K$ can be selected arbitrarily large, and $L$ is fixed and finite satisfying $L \geq \max\left\{L_{corr}, L_{isi}\right\}$, then, by letting $K \rightarrow  \infty$, the rate loss due to discarding $L - 1$ samples out of every $N^{(K)}$ samples approaches zero, and the capacity of the transformed MIMO channel asymptotically corresponds to the exact capacity of the original scalar channel with ACGN, see, e.g., \cite[Pg. 181]{Tse:05} and \cite[Sec. III]{Dabora:13}.
Mathematically, let $I(\cdot ; \cdot)$ and $p(\cdot)$ denote the mutual information and the probability density function, respectively. Also, for any sequence $q[n]$, $n \in \mathds{Z}$, and integers $a_1 < a_2$, we use ${\bf q}_{a_1}^{a_2}$ to denote the column vector $\left[q[a_1],\ldots,q[a_2 -1]\right]^T$ and ${\bf q}^{a_2}$ to denote ${\bf q}_{0}^{a_2}$.
Let ${\bf S}_0 \in \mathcal{S}$ denote the initial state of the channel, i.e., ${\bf S}_0 = \left[\left({\bf x}_{-L+1}^0\right)^T,\left({\bf w}_{-L+1}^0\right)^T\right]^T$, $\mathcal{S} = \mathds{R}^{2(L-1)}$.
Lastly, define $\beta_N \triangleq \frac{1}{N}\sum\limits_{m=0}^{N-1}\E\{|x[m]|^2\}$, and
\vspace{-0.2cm}
\begin{equation*}
\label{eqn:Sec_DefCn}
{C_k}\left( {{{{\bf s} }_0}} \right) \triangleq \frac{1}{k}\mathop {\sup }\limits_{p\left( {{{\bf x} }^k} \right):{\beta_k} \le \rho } I\left( {\left. {{{{\bf x} }^k};{{{\bf r} }^k}} \right|{{{\bf S} }_0} = {{{\bf s} }_0}} \right).
\end{equation*}
%
The proof combines elements from the capacity analysis of finite memory point-to-point channels in \cite{Massey:88}, of multiaccess channels in \cite{Verdu:89}, and of broadcast channels in \cite{Goldsmith:01}.
{\bf A complete detailed proof is provided in Appendix \ref{app:Proof1a}}.
In the following lemmas we provide upper and lower bounds to the capacity of the channel \eqref{eqn:ChannelModel1}:

\begin{lemma}
\label{lem:Asym_Eq_0}
For every $\epsilon \in (0,1)$, $\gamma > 0$, $\exists K_0 >0$ such that $\forall k > K_0$, $\forall {\bf s}_0 \in \mathcal{S}$,
the capacity of the channel \eqref{eqn:ChannelModel1} satisfies
\begin{equation*}
C_{LPTV-ACGN}  \le \frac{1}{1 - {\epsilon }}\mathop {\inf }\limits_{{{{\bf s} }_0} \in \mathcal{S}}C_k\left({\bf s}_0\right) + \frac{1}{\left({1 - {\epsilon }}\right)k} + \gamma.
\end{equation*}
\end{lemma}
{\em Proof:} Similarly to the derivation in \cite[Eq. (5)-(8)]{Verdu:89}, it can be shown that for any arbitrary initial condition ${\bf s}_0$, for all $\epsilon, \gamma > 0$ and sufficiently large positive integer $k$, every rate $R$ achievable for the channel \eqref{eqn:ChannelModel1} satisfies
\begin{equation}
\label{eqn:app_proof4}
\left( {1 - {\epsilon }} \right)\left( {{R} - {\gamma}} \right) - \frac{1}{k} \leq {C_k}\left( {{{{\bf s} }_0}}\right).
\end{equation}
Since \eqref{eqn:app_proof4} holds for all sufficiently large $k$, for $\epsilon < 1$, by dividing both sides of \eqref{eqn:app_proof4} by $1 - \epsilon$ the lemma follows.
Lemma \ref{lem:Asym_Eq_0} is restated as Lemma \ref{Cor:Proof_C2} in Appendix \ref{app:Proof1a} where a detailed proof is provided.
%
%
$\blacksquare$
\smallskip

\begin{lemma}
\label{lem:Asym_Eq_1}
$\mathop { \inf }\limits_{{\bf s}_0 \in \mathcal{S}}C_{N^{(K)}}\left( {{{{\bf s} }_0}} \right) \leq \frac{K+K_{min}}{K}R_{K+K_{min}}$.
\end{lemma}
{\em Proof:} First, recall that $R_K$ denotes the capacity of the transformed memoryless $N^{(K)}\times M^{(K)}$ MIMO channel \eqref{eqn:Cap_LPTV4}. It therefore follows that $R_K$ is independent the initial state and can also be written as \cite[Ch. 9.1]{ElGamal:10}
\begin{equation*}
R_K = \frac{1}{N^{(K)}}\mathop {\sup }\limits_{p\left( {{{{\bf x} }^{N^{(K)}}}} \right):{\beta_{N^{(K)}}} \le \rho } I\left( {{{{\bf x} }^{N^{(K)}}};{\bf r} _{L-1}^{N^{(K)}}} \right).
\end{equation*}
Next, similarly to the derivation leading to \cite[Eq. (32)]{Massey:88}, it is shown that $\mathop { \inf }\limits_{{\bf s}_0 \in \mathcal{S}}N^{(K)}C_{N^{(K)}}\left( {{{{\bf s} }_0}} \right) \le N^{(K+K_{min})}R_{K+K_{min}}$. By dividing both sides by $N^{(K)}$, the lemma follows.
Lemma \ref{lem:Asym_Eq_1} is restated as Lemma \ref{lem:proof2} in Appendix \ref{app:Proof1a} where a detailed proof is provided.
$\blacksquare$
\smallskip

\begin{lemma}
\label{lem:Asym_Eq_2}
$\mathop {\sup }\limits_{K > K_{min}} R_K \le C_{LPTV-ACGN}$.
\end{lemma}
{\em Proof:} This lemma is a special case of the more general inequality proved for the multiterminal channel in \cite[Lemma 2]{Goldsmith:01}. From  \cite[Lemma 2]{Goldsmith:01} it follows that any code for the equivalent channel in which $L-1$ channel outputs out of every $N^{(K)}$ channel outputs are discarded, can be applied to the original channel \eqref{eqn:ChannelModel1} s.t. the rate and probability of error are maintained. This is done by using the same encoding and decoding scheme, where only the last $M^{(K)}$ out of every $N^{(K)}$ channel outputs are employed in  decoding.
It therefore follows that any rate achievable for the transformed channel is also achievable for the original channel \eqref{eqn:ChannelModel1}.
Lemma \ref{lem:Asym_Eq_2} is restated as Lemma \ref{lem:proof3} in Appendix \ref{app:Proof1a} where a detailed proof is provided.
$\blacksquare$
\smallskip

From the above lemmas we now conclude that for every $\epsilon \in (0,1)$, $\gamma > 0$, $\exists K_0 >0$ such that $\forall K > K_0$
\begin{align}
\mathop {\sup }\limits_{{\tilde{K}} > K_{min}} R_{\tilde{K}}
&\stackrel{(a)}{\leq} C_{LPTV-ACGN} \notag \\
&\stackrel{(b)}{\leq} \frac{1}{1 - {\epsilon }}\mathop {\inf }\limits_{{{{\bf s} }_0} \in \mathcal{S}}C_{N^{(K)}}\left({\bf s}_0\right) + \frac{1}{\left({1 - {\epsilon }}\right){N^{(K)}}} + \gamma \notag  \\
&\stackrel{(c)}{\leq} \frac{K+K_{min}}{\left({1 - {\epsilon }}\right)K}R_{K+K_{min}} + \frac{1}{\left({1 - {\epsilon }}\right){N^{(K)}}} + \gamma, \notag 
\end{align}
where $(a)$ follows from Lemma \ref{lem:Asym_Eq_2}; $(b)$ follows from Lemma \ref{lem:Asym_Eq_0}; $(c)$ follows from Lemma \ref{lem:Asym_Eq_1}.
As this is satisfied for all sufficiently large $K$, it follows that
\begin{align}
C_{LPTV-ACGN}
& \leq \mathop {\lim \inf }\limits_{K \to \infty } \left(\frac{K+K_{min}}{\left({1 - {\epsilon }}\right)K}R_{{K+K_{min}}} +  \frac{1}{\left({1 - {\epsilon }}\right){N^{(K)}}}\right) + \gamma \notag \\
& =  \frac{1}{1 - {\epsilon }}\mathop {\lim \inf }\limits_{K \to \infty } R_K  + \gamma. \label{eqn:Last2}
\end{align}
Since  $\epsilon , \gamma$ can be made arbitrarily small, \eqref{eqn:Last2} implies that
\begin{equation*}
C_{LPTV-ACGN} \leq \mathop {\lim \inf }\limits_{K \to \infty } R_K.
\end{equation*}
Lastly, it follows from the definition of $\lim \sup$ \cite[Def. 5.4]{Amann:05} that $\mathop {\lim \sup }\limits_{{K} \to \infty }R_{{K}} \leq \mathop {\sup }\limits_{{\tilde{K}} > K_{min}} R_{\tilde{K}}$. Since $\mathop {\lim \inf }\limits_{K \to \infty } R_K \leq \mathop {\lim \sup }\limits_{K \to \infty } R_K$, it follows that
\begin{equation*}
C_{LPTV-ACGN} = \mathop {\lim }\limits_{K \to \infty }R_K,
\end{equation*}
which completes the proof that \eqref{eqn:SISO_Cap7} denotes the capacity of the LPTV channel with  channel for $K \rightarrow \infty$.
$\blacksquare$


\vspace{-0.2cm}
\chapter{Detailed Proof of Asymptotic Equivalence for Thm. \ref{thm:CapacityLPTV}}
\label{app:Proof1a}
\vspace{-0.1cm}

Appendix \ref{app:Proof1} details the derivation of the capacity of DT LPTV channels with ACGN. The derivation is based on the analysis of the capacity of an equivalent channel, in which at each block a finite and fixed number of channel outputs are discarded, and the blocklength is an integer multiple of the least common multiple of the periods of the channel and the noise. The capacity is obtained by taking the limit of the blocklength to infinity, referred to in the following as the {\em asymptotic blocklength}.
In this appendix we prove that at the asymptotic blocklength, the capacity of the equivalent channel is equal to the capacity of the original LPTV ACGN channel.
Our derivation here follow similar proofs in \cite[Sec. IV]{Massey:88}, \cite[Sec. II]{Verdu:89}, and \cite[Appendix A]{Goldsmith:01}, where the main difference follow as all these works considered LTI channels with colored stationary Gaussian noise, while we consider LPTV channels with ACGN.
We use $I(X;Y)$ to denote the mutual information between the RVs $X \in \mathcal{X}$ and $Y \in \mathcal{Y}$, $H(X)$ to denote the entropy of $X$, $p(X)$ to denote the probability density function (PDF) of $X$, and $p_X(x)$ to denote the PDF evaluated at $x$. 
For any sequence, possibly multivariate, ${\bf q}[m]$, $m \in \mathds{Z}$, and integers $a_1 < a_2$, we use ${\bf q}_{a_1}^{a_2}$ to denote the column vector $\left[{\bf q}[a_1]^T,\ldots,{\bf q}[a_2-1]^T\right]^T$ and ${\bf q}^{a_2}$ to denote ${\bf q}_{0}^{a_2}$.

\section{Definitions}
We begin with establishing the definitions for a channel, a channel code, achievable rate, memoryless channels, and block memoryless channels.

\begin{definition}
\label{def:Channel}
A channel consists of a (possibly multivariate) input stream  ${\bf{x}}[m] \in  \mathcal{X} \triangleq \mathds{R}^{n_x}$, a (possibly multivariate) output streams  ${\bf{y}}[m] \in \mathcal{Y} \triangleq \mathds{R}^{n_y}$, $m \in \mathds{N}$, an initial state ${\bf S}_0 \in \mathcal{S} \triangleq \mathds{R}^{n_s}$, $n_x, n_y, n_s \in \mathds{N}$ fixed, and a sequence of transition probabilities $\left\{p\left({\bf{y}}^t|{\bf{x}}^t,{\bf S}_0\right)\right\}_{t=0}^{\infty}$, such that for all $t \in\mathds{N}$, ${\bf{\beta}}^t \in \mathcal{X}^t$, ${\bf s}_0 \in \mathcal{S}$
\begin{equation*}
\int\limits_{\mathcal{Y}^t} p_{{\bf{y}}^t|{\bf{x}}^t,{\bf S}_0}\left({\bf{\alpha}}^t|{\bf{\beta}}^t,{\bf s}_0\right) d{\bf{\alpha}}^t = 1.
\end{equation*}
\end{definition}
The observer of channel output $\mathcal{Y}$ is referred to as the {\em receiver}.

\begin{definition}
\label{def:Code}
Let $n$ and $R$ be positive integers. A $\left[R, n \right]$ code consists of an encoder $e_n$ which maps a message $M$ uniformly distributed on $\mathcal{M} \triangleq \{0,1,\ldots,2^{nR}-1\}$ into a codeword ${\bf{x}}^n \in\mathcal{X}^n$, i.e.,
\begin{equation*}
e_n:\mathcal{M} \mapsto \mathcal{X}^n,
\end{equation*}
and a decoder $d_n$ which maps the channel output  ${\bf{y}}^n \in\mathcal{Y}^n$ into a message word $\hat{M}$, i.e.,
\begin{equation*}
d_n: \mathcal{Y}^n \mapsto \mathcal{M}.
\end{equation*}
The encoder and the message are assumed to be independent of the initial state ${\bf s}_0$.
\end{definition}
%
Since the message is uniformly distributed, the average probability of error is given by
\begin{equation}
\label{eqn:def_avgError}
P_e^{(n)} \left( {{\bf s} }_0 \right) = \frac{1}{2^{nR} }\sum\limits_{m = 0}^{2^{nR}  - 1} \Pr \left( {\left. {{d_n }\left( {\bf y}^n \right) \ne m} \right|M = m,{{{\bf S} }_0} = {{{\bf s} }_0}} \right)
\end{equation}
%
\begin{definition}
\label{def:SecrecyRate}
A  rate $R$ is achievable for a channel if for every $\epsilon,\gamma > 0$, $\exists n_0 >0$ such that $\forall n > n_0$ there exists a $\left[R_1, n \right]$ code which satisfies
\begin{subequations}
\label{eqn:def_Rs}
\begin{equation}
\label{eqn:def_Rs1}
\mathop {\sup }\limits_{{{{\bf s} }_0} \in \mathcal{S}} P_e^{(n)} \left( {{{{\bf s} }_0}} \right) < {\epsilon },
\end{equation}
and
\begin{equation}
\label{eqn:def_Rs3}
R_1 \geq R - \gamma.
\end{equation}
\end{subequations}
\end{definition}
The supremum of all achievable rates is called the {\em channel capacity}.

\begin{definition}
\label{def:MemorylessChannel}
A channel  is said to be {memoryless} if for every positive integer $k$
\begin{equation*}
p\left({\bf{y}}^k|{\bf{x}}^k,{\bf S}_0\right) = \prod\limits_{m=0}^{k-1}p\left({\bf{y}}[m]|{\bf{x}}[m]\right).
\end{equation*}
\end{definition}
\begin{definition}
\label{def:BlockMemorylessChannel}
A channel is said to be {$t$-block memoryless} if for every positive integer $k$
\begin{equation*}
p\left({\bf{y}}^{t\cdot k}|{\bf{x}}^{t\cdot k},{\bf S}_0\right) = \prod\limits_{l=1}^{k-1}p\left({\bf{y}}_{t\cdot (l-1)}^{t\cdot l}|{\bf{x}}_{t\cdot (l-1)}^{t\cdot l}\right).
\end{equation*}
\end{definition}
Note that the average error probability $P_e^{(n)} $ is independent of the initial state ${\bf S}_0$ for memoryless channel and for $t$-block memoryless channel when $n$ is an integer multiple of $t$.

\section{Channel Models}
We consider a scalar passband DT LPTV channel with ACGN. Let $w[m]$ denote the ACGN with period $N_{noise}$ and finite memory $L_{corr}$, i.e., $c_{ww}(m,l) \triangleq \E\{w[m+l]w[m]\} = c_{ww}(m+N_{noise},l)$, $\forall m,l \in \mathds{Z}$, and $c_{ww}(m,l) = 0$, $\forall |l| \geq L_{corr}$. 
Let $g[m,l]$ denote the channel impulse response, whose memory is denoted by $L_{isi}$ and period is denoted by $N_{ch}$, i.e., $g[m,l]= g\left[m+N_{ch},l\right]$, $\forall m,l \in \mathds{Z}$, and $g[m,l] = 0$, $\forall |l| \geq L_{isi}$.
Let $x[m]$ denote the channel input and $r[m]$ denote the channel output, the input-output relationship of the channel is given by
\begin{equation}
\label{eqn:model_LPGC1}
r[m]=\sum\limits_{l=0}^{L_{isi}-1}g[m,l]x[m-l] + w[m],
\end{equation}
$m \in \{0,1,\ldots,N\}$.
For any integers $a_1 < a_2$, define  $\beta_{a_1}^{a_2} \triangleq \frac{1}{a_2-a_1}\sum\limits_{m=a_1}^{a_2-1}\E\left\{|x[m]|^2\right\}$ and  $\beta_{a_2} \triangleq \beta_{0}^{a_2}$.
We consider an average power constraint on the channel input
\begin{equation}
\label{eqn:model_LPGC2}
\beta_N =\frac{1}{N}\sum\limits_{m=0}^{N-1}\E\left\{|x[m]|^2\right\} \leq \rho.
\end{equation}
In the following we refer to this channel as the linear periodic Gaussian channel (LPGC).
We use $N_{lcm}$ to denote the least common multiple of $N_{noise}$ and $N_{ch}$. 
Define $L \triangleq \max\left\{L_{isi}, L_{corr}\right\}$, $N^{(K)} \triangleq K N_{lcm}$, and $M^{(K)} \triangleq N^{(K)} - L + 1$.
Note that the LPGC is not memoryless, and that the initial state of this channel is ${\bf S}_0 = \left[\left({\bf x}_{-L+1}^0\right)^T,\left({\bf w}_{-L+1}^0\right)^T\right]^T \in \mathcal{S} \triangleq \mathds{R}^{2(L-1)}$.

In the sequel we analyze the asymptotic expression for the capacity of the LPGC.
Similarly to the derivations of the capacity of the finite-memory Gaussian channels for point-to-point communications \cite{Massey:88}, multiaccess communications \cite{Verdu:89}, and broadcast communications \cite{Goldsmith:01}, for $t > L$ we define the $t$-block memoryless periodic Gaussian channel ($t$-MPGC), which is obtained from the LPGC by considering the last $t-L+1$ channel outputs over each $t$-block, i.e., the outputs of the $t$-MPGC are defined as the outputs of the LPGC for $n\% t \geq L-1$, while for  $n\% t < L-1$ the outputs of the $t$-MPGC are undefined.
The $t$-MPGC inherits the power constraints of the LPGC \eqref{eqn:model_LPGC2}.

Note that the LPGC corresponds to the original LPTV channel with ACGN model \eqref{eqn:ChannelModel1}, while the $t$-MPGC, with $t = N^{(K)}$, corresponds to the equivalent channel used for obtaining the time-invariant MIMO model in the proof of Thm. \ref{thm:CapacityLPTV} in Appendix \ref{app:Proof1}. Therefore, $C_{LPTV-ACGN}$ and $R_K$ are the capacity of the LGPC and the capacity of the $N^{(K)}$-MPGC, respectively.

\section{Equivalence of the Capacity at the Asymptotic Blocklength}
In the following we prove that $C_{LPTV-ACGN} = \mathop {\lim }\limits_{K \to \infty } R_K$.
We begin by proving the following:
%
\begin{proposition}
\label{lem:proof1}
The capacity of the $N^{(K)}$-MPGC is given by
\begin{equation}
\label{eqn:proof1}
R_K = \frac{1}{N^{(K)}} \mathop {\sup }\limits_{p\left( {\bf x}^{N^{(K)}} \right):{\beta_{N^{(K)}}} \le \rho } I\left( {\bf x}^{N^{(K)}};{\bf r}_{L-1}^{N^{(K)}} \right).
\end{equation}
\end{proposition}
{\em Proof:}
In order to obtain the capacity of the ${N^{(K)}}$-MPGC, we first show that \eqref{eqn:proof1} denotes the maximum achievable rate when considering only codes whose length is an integer multiple of ${N^{(K)}}$, i.e, $\left[R_1, b \cdot {N^{(K)}}\right]$ codes where $b \in \mathds{N}$. Then, we show that every rate achievable for the ${N^{(K)}}$-MPGC can be achieved by considering only codes whose length is an integer multiple of ${N^{(K)}}$.

Let us consider the ${N^{(K)}}$-MPGC subject to the limitation that only codes whose length is an integer multiple of ${N^{(K)}}$ are allowed. In this case we can transform the channel into an equivalent ${N^{(K)}}  \times  {M^{(K)}}$ memoryless MIMO channel without loss of information as was done in Appendix \ref{app:Proof1}. We hereby repeat this transformation with a slight change of notations compared to that of Appendix \ref{app:Proof1} in order to maintain consistency throughout this appendix. These changes are clearly highlighted in the following.
Define the input of the transformed channel by the ${N^{(K)}} \times 1$ vector ${{\bf x} _{eq}}\left[\tilde n\right] \triangleq {\bf x} _{\left( {\tilde n - 1} \right) \cdot {N^{(K)}}}^{\tilde n \cdot {N^{(K)}}}$ (corresponds to ${\bf x}[n]$ in Appendix \ref{app:Proof1}), and the output of the transformed channel by the ${M^{(K)}} \times 1$ vector ${{\bf r} _{eq}}\left[\tilde n\right] \triangleq {\bf r} _{\left( {\tilde n - 1} \right) \cdot {N^{(K)}} + L-1}^{\tilde n \cdot {N^{(K)}}}$ (corresponds to ${\bf r}[n]$ in Appendix \ref{app:Proof1}).
The transformation is clearly reversible thus the capacity of the transformed channel is equal to the capacity of the original channel.
Since the ${N^{(K)}}$-MPGC is ${N^{(K)}}$-block memoryless, it follows from Definition \ref{def:MemorylessChannel} that the transformed MIMO channel is memoryless. The channel output is corrupted by the additive noise vectors ${{\bf w} _{eq}}\left[\tilde n\right] \triangleq {\bf w} _{\left( {\tilde n - 1} \right) \cdot {N^{(K)}} + L-1}^{\tilde n \cdot {N^{(K)}}}$ (corresponds to ${\bf w}[n]$ in Appendix \ref{app:Proof1}). From definition it follows that ${{\bf w} _{eq}}\left[\tilde n\right]$ is a zero-mean multivariate Gaussian process. From the properties of the DCD (see Lemma \ref{pro:NoiseCorr}) and the memoryless property of the transformed channel (see discussion following Eq. \eqref{eqn:Cap_LPTV3} in Appendix \ref{app:Proof1}), it follows that  ${{\bf w} _{eq}}\left[\tilde n\right]$ is i.i.d. over $\tilde n$.
Since we assume codewords of length $b \cdot {N^{(K)}}$, the average power constraint \eqref{eqn:model_LPGC2} implies that
\begin{equation}
\label{eqn:Constraint2b}
\E\left\{ \frac{1}{b}\sum\limits_{\tilde n = 0}^{b - 1} {\left\| {{{{\bf x} }_{eq}}\left[ {\tilde n} \right]} \right\|}^2  \right\} = \frac{1}{b\cdot {N^{(K)}}}\sum\limits_{m = 0}^{b \cdot N^{(K)} - 1}\E\left\{  {{\left| x\left[ m \right] \right|^2}}  \right\} \cdot {N^{(K)}} \le \rho  \cdot {N^{(K)}}.
\end{equation}
The capacity of the transformed channel with the constraint \eqref{eqn:Constraint2b}, denoted $C_{MIMO}^{(K)}$, is well-established, and is given by  \cite[Ch. 10.3]{Goldsmith:05}:
\begin{equation}
\label{eqn:EqSecCap1}
C_{MIMO}^{(K)} = \mathop {\sup }\limits_{p\left( {{{{\bf x} }^{N^{(K)}}}} \right):{\beta_{N^{(K)}}} \le \rho } I\left( {{{{\bf x} }^{N^{(K)}}};{\bf r} _{L-1}^{N^{(K)}}} \right).
\end{equation}
%
As each MIMO channel use corresponds to ${N^{(K)}}$ channel uses in the original channel, it follows that the achievable rate of the ${N^{(K)}}$-MPGC subject to the limitation that only codes whose length is an integer multiple of ${N^{(K)}}$ are allowed, in bits per channel uses, is $\frac{1}{N^{(K)}}C_{MIMO}^{(K)}$, which coincides with \eqref{eqn:proof1}.

Next, we show that any rate achievable for the $N^{(K)}$-MPGC can be achieved by considering only codes whose length is an integer multiple of $N^{(K)}$. Consider a rate $R$ achievable for the $N^{(K)}$-MPGC and fix $\epsilon , \gamma $. From Definition \ref{def:SecrecyRate}  it follows that $\exists n _0 > 0$ such that $\forall n > n _0$ there exists a $\left[R_1, n\right]$ code which satisfies \eqref{eqn:def_Rs1}-\eqref{eqn:def_Rs3}. Thus, by setting $b_0$ as the smallest integer $b_0$ for which $b_0 \cdot {N^{(K)}} \geq n _0$ it follows that for all integer $b > b_0$ there exists a $\left[R_1 , b \cdot N^{(K)}\right]$ code which satisfies \eqref{eqn:def_Rs1}-\eqref{eqn:def_Rs3}. Therefore, the rate $R$ is also achievable when considering only codes whose blocklength is an integer multiple of ${N^{(K)}}$. We therefore conclude that $R_K$ denotes the maximum achievable rate for the ${N^{(K)}}$-MPGC, which proves the proposition.
$\blacksquare$
\smallskip

\begin{proposition}
\label{Cor:Proof_Qn2}
Proposition \ref{lem:proof1} implies that for any arbitrary initial condition ${{\bf s} }_0$
\begin{equation}
\label{eqn:Cor_Qn2}
R_K = \frac{1}{N^{(K)}} \mathop {\sup }\limits_{p\left( {\bf x}^{N^{(K)}} \right):{\beta_{N^{(K)}}} \le \rho } I\left( {\bf x}^{N^{(K)}};{\bf r}_{L-1}^{N^{(K)}}| {{\bf S} }_0 = {{\bf s} }_0 \right).
\end{equation}
\end{proposition}

{\em Proof:} Note that $C_{MIMO}^{(K)}$ denotes the maximum achievable rate of an ${N^{(K)}}$-block memoryless channel when considering only blocklength which are a multiple of ${N^{(K)}}$. Thus, from Definition \ref{def:BlockMemorylessChannel} it follows that  $C_{MIMO}^{(K)}$ is independent of the initial state.
Since $R_K = \frac{1}{N^{(K)}}C_{MIMO}^{(K)}$, it follows that $R_K$ is also independent of the initial state, hence \eqref{eqn:Cor_Qn2} follows from \eqref{eqn:proof1}.
$\blacksquare$
\smallskip

Before we proceed, let us recall the definition of ${C_k}\left( {{{{\bf s} }_0}} \right)$:
\begin{equation*}
{C_k}\left( {{{{\bf s} }_0}} \right) \triangleq \frac{1}{k}\mathop {\sup }\limits_{p\left( {{{\bf x} }^k} \right):{\beta _k} \le \rho } I\left( {\left. {{{{\bf x} }^k};{{{\bf r} }^k}} \right|{{{\bf S} }_0} = {{{\bf s} }_0}} \right),
\end{equation*}

\begin{proposition}
\label{lem:Proof_Thm1a}
$C_{LPTV-ACGN} \le \mathop {\inf }\limits_{{{{\bf s} }_0} \in \mathcal{S}} \left( {\mathop {\lim \inf }\limits_{k \to \infty } {C_k}\left( {{{{\bf s} }_0}} \right)} \right)$.
\end{proposition}
{\em Proof:} We prove the proposition by showing that every rate $R$ achievable for the LPGC satisfies $R \le \mathop {\lim \inf }\limits_{k \to \infty } {C_k}\left( {{{{\bf s} }_0}} \right)$ for any initial condition ${{{{\bf s} }_0}}$. By definition, if $R$ is achievable then for every $\epsilon , \gamma > 0$ and for all sufficiently large $n$ there exists a $[R_1, n]$ code, i.e., a code with blocklength $n$ and a message $M$ uniformly distributed over $\mathcal{M}$, such that \eqref{eqn:def_Rs1}-\eqref{eqn:def_Rs3} are satisfied. Fix an initial condition ${{\bf s}}_0$, since conditioning only reduces entropy it follows that
\begin{align}
H\left( {\left. M \right|{{{\bf r} }^n },{{{\bf S} }_0} = {{{\bf{ s}} }_0}} \right)
&\stackrel{(a)}{\leq} 1 + \Pr \left( \left.{M \ne \hat M} \right|{\bf S}_0 = {\bf s}_0 \right) n R_1\notag \\
&\stackrel{(b)}{\leq} 1 + {\epsilon} \cdot n R_1, \label{eqn:app_proof1}
\end{align}
where $(a)$ follows from Fano's inequality \cite[Sec. 2.10]{Cover:00} and $(b)$ follows from \eqref{eqn:def_Rs1} since
\begin{equation*}
\Pr \left( \left.{M \ne \hat M} \right|{\bf S}_0 = {\bf s}_0 \right) \le \mathop {\sup }\limits_{{{{\bf s} }_0}} \Pr\Big(  M \ne \hat M \big|{{{\bf S} }_0} = {{{\bf s} }_0} \Big) \le {\epsilon}.
\end{equation*}
 Therefore,
\begin{align}
I\left( {\left. {M;{{{\bf r} }^n }} \right|{{{\bf S} }_0} = {{{\bf{ s}} }_0}} \right)
&= H\left( {\left. M \right|{{{\bf S} }_0} = {{{\bf{ s}} }_0}} \right) - H\left( {\left. M \right|{{{\bf r} }^n},{{{\bf S} }_0} = {{{\bf{ s}} }_0}} \right) \notag \\
&\stackrel{(a)}{\geq} H\left( {\left. M \right|{{{\bf S} }_0} = {{{\bf{ s}} }_0}} \right) - 1 -  {\epsilon} \cdot n R_1 \notag \\
&\stackrel{(b)}{=}n R_1 - 1 -  {\epsilon} \cdot n R_1, \label{eqn:app_proof2}
\end{align}
where $(a)$ follows from \eqref{eqn:app_proof1}, and $(c)$ follows since $M$ is uniformly distributed and independent of ${{\bf s}}_0$, thus $H\left( {\left. M \right|{{{\bf S} }_0} = {{{\bf{ s}} }_0}} \right) = H\left( M \right) = n R_1$.
Combining \eqref{eqn:def_Rs3} and \eqref{eqn:app_proof2} leads to
\begin{equation*}
{R} - {\gamma}
 \le \frac{{I\left( {\left. {M;{{{\bf r} }^n }} \right|{{{\bf S} }_0} = {{{\bf{ s}} }_0}} \right)  + 1 }}{{n \left( {1 - {\epsilon }} \right)}},
\end{equation*}
thus
\begin{align}
\left( {1 - {\epsilon }} \right)\left( {{R} - {\gamma}} \right) - \frac{1}{n }
&\le \frac{1}{n }I\left( {\left. {M;{{{\bf r} }^n }} \right|{{{\bf S} }_0} = {{{\bf{ s}} }_0}} \right) \notag \\
&\stackrel{(a)}{\leq} \frac{1}{n }\mathop {\sup }\limits_{p\left( {{{\bf x} }^n } \right):{\beta _{n}} \le \rho } I\left( {\left. {{{{\bf x} }^n };{{{\bf r} }^n }} \right|{{{\bf S} }_0} = {{{\bf{ s}} }_0}} \right) \notag \\
&= {C_n }\left( {{{{\bf{ s}} }_0}} \right), \label{eqn:app_proof3}
\end{align}
where $(a)$ follows from the data-processing lemma \cite[Sec. 2.8]{Cover:00} as $M|{\bf S}_0 \rightarrow  {{{\bf x} }^n }|{\bf S}_0  \rightarrow  {{{\bf r} }^n }|{\bf S}_0 $ form a Markov chain.
Since \eqref{eqn:app_proof3} holds for all sufficiently large $n$, it follows that
\begin{equation}
\label{eqn:app_proof4b}
\left( {1 - {\epsilon }} \right)\left( {{R} - {\gamma}} \right) \le \mathop {\lim \inf }\limits_{n \to \infty } {C_n }\left( {{{{\bf{ s}} }_0}} \right).
\end{equation}
Since  $\epsilon , \gamma$ can be made arbitrarily small, \eqref{eqn:app_proof4b} implies that for all ${{{{\bf{ s}} }_0}}$, ${R} \le \mathop {\lim \inf }\limits_{n  \to \infty } {C_n }\left( {{{{\bf{s}} }_0}} \right)$.
$\blacksquare$
\smallskip

We now repeat the lemmas stated in Appendix \ref{app:Proof1} and provide a detailed proof for each lemma.
\begin{lemma}
\label{Cor:Proof_C2}
For every $\epsilon \in (0,1)$, $\gamma > 0$, $\exists N >0$ such that $\forall n > N$
\begin{equation}
\label{eqn:Cor_C2}
C_{LPTV-ACGN} \leq \frac{1}{1 - {\epsilon }}\mathop {\inf }\limits_{{{{\bf s} }_0} \in \mathcal{S}}C_n\left({\bf s}_0\right) + \frac{1}{\left({1 - {\epsilon }}\right)n} + \gamma.
\end{equation}
\end{lemma}

{\em Proof:} This lemma follows immediately from Proposition \ref{lem:Proof_Thm1a}, as \eqref{eqn:app_proof3} is satisfied for all rates achievable for the LPGC, for all initial states ${{{{\bf{ s}} }_0}}$. As $\epsilon < 1$, dividing both sides of \eqref{eqn:app_proof3} by $1 - \epsilon$ leads to \eqref{eqn:Cor_C2}.
$\blacksquare$
\smallskip

\begin{lemma}
\label{lem:proof2}
$\mathop { \inf }\limits_{{\bf s}_0 \in \mathcal{S}}C_{N^{(K)}}\left( {{{{\bf s} }_0}} \right) \leq \frac{K+K_{min}}{K}R_{K+K_{min}}$.
\end{lemma}
{\em Proof:}
Note that
\begin{align}
&{N^{(K \!+ \! {K_{min}})}}{R_{{{K \!+ \! {K_{min}}}}}} \notag \\
&\quad\! = \! \mathop {\sup }\limits_{p\left( {{{\bf{x}}^{{N^{\left( {K \!+ \! {K_{min}}} \right)}}}}} \right):~{\beta _{{N^{\left( {K \!+ \! {K_{min}}} \right)}}}} \le \rho } I\left( {{{\bf{x}}^{{N^{\left( {K \!+ \! {K_{min}}} \right)}}}};{\bf{r}}_{L \! - \! 1}^{{N^{\left( {K \!+ \! {K_{min}}} \right)}}}} \right) \notag \\
&\quad\stackrel{(a)}{\geq} \mathop {\sup }\limits_{p\left( {{{\bf{x}}^{{N^{\left( {K \!+ \! {K_{min}}} \right)}}}}} \right):  ~\beta _{{N^{\left( {K_{min}} \right)}}}^{{N^{\left( {K \!+ \! {K_{min}}} \right)}}} \le \rho ; ~{\beta _{{N^{\left( {K_{min}} \right)}}}} \le \rho} I\left( {\left. {{\bf{x}}_{{N^{\left( {{K_{min}}} \right)}}}^{{N^{\left( {K \!+ \! {K_{min}}} \right)}}};{\bf{r}}_{{N^{\left( {{K_{min}}} \right)}}}^{{N^{\left( {K \!+ \! {K_{min}}} \right)}}}} \right|{{\bf{x}}^{{N^{\left( {{K_{min}}} \right)}}}},{\bf{r}}_{L \! - \! 1}^{{N^{\left( {{K_{min}}} \right)}}}} \right) \notag \\
&\quad\stackrel{(b)}{\! = \!} \mathop {\sup }\limits_{p\left( {{{\bf{x}}^{{N^{\left( {K \!+ \! {K_{min}}} \right)}}}}} \right):  ~\beta _{{N^{\left( {K_{min}} \right)}}}^{{N^{\left( {K \!+ \! {K_{min}}} \right)}}} \le \rho ; ~{\beta _{{N^{\left( {K_{min}} \right)}}}} \le \rho} I\left( {\left. {{\bf{x}}_{{N^{\left( {{K_{min}}} \right)}}}^{{N^{\left( {K \!+ \! {K_{min}}} \right)}}};{\bf{r}}_{{N^{\left( {{K_{min}}} \right)}}}^{{N^{\left( {K \!+ \! {K_{min}}} \right)}}}} \right|{\bf{x}}_{{N^{\left( {{K_{min}}} \right)}} \! - \! L \!+ \! 1}^{{N^{\left( {{K_{min}}} \right)}}},{\bf{w}}_{{N^{\left( {{K_{min}}} \right)}} \! - \! L \!+ \! 1}^{{N^{\left( {{K_{min}}} \right)}}}} \right) \notag \\
&\quad\stackrel{(c)}{\! = \!} \mathop {\sup }\limits_{p\left( {{{\bf{x}}^{{N^{\left( {K \!+ \! {K_{min}}} \right)}}}}} \right):  ~\beta _{{N^{\left( {K_{min}} \right)}}}^{{N^{\left( {K \!+ \! {K_{min}}} \right)}}} \le \rho ; ~{\beta _{{N^{\left( {K_{min}} \right)}}}} \le \rho} ~\int\limits_{{{\bf{\tilde s}}_0} \in \mathcal{S}} I\Big(\left. {{\bf{x}}_{{N^{\left( {{K_{min}}} \right)}}}^{{N^{\left( {K \!+ \! {K_{min}}} \right)}}};{\bf{r}}_{{N^{\left( {{K_{min}}} \right)}}}^{{N^{\left( {K \!+ \! {K_{min}}} \right)}}}} \right| \notag \\
& \qquad \qquad \qquad \qquad \qquad \qquad \qquad  \qquad \qquad  \qquad \quad \left\{ {{\bf{x}}_{{N^{\left( {{K_{min}}} \right)}} \! - \! L \!+ \! 1}^{{N^{\left( {{K_{min}}} \right)}}},{\bf{w}}_{{N^{\left( {{K_{min}}} \right)}} \! - \! L \!+ \! 1}^{{N^{\left( {{K_{min}}} \right)}}}} \right\} \! = \! {{\bf{\tilde  s}}_0} \Big) \notag \\
& \qquad \qquad \qquad \qquad \qquad \qquad \qquad \qquad  \qquad  \quad \times {p_{{{\bf{x}}_{{N^{\left( {{K_{min}}} \right)}} \! - \! L \!+ \! 1}^{{N^{\left( {{K_{min}}} \right)}}},{\bf{w}}_{{N^{\left( {{K_{min}}} \right)}} \! - \! L \!+ \! 1}^{{N^{\left( {{K_{min}}} \right)}}}}}}\left( {{{\bf{\tilde  s}}_0}} \right)d{{\bf{\tilde s}}_0} \notag \\
&\quad\stackrel{(d)}{\geq} \mathop {\inf }\limits_{{{\bf{s}}_0} \in \mathcal{S}} \mathop {\sup }\limits_{p\left( {{\bf{x}}_{{N^{({K_{min}})}}}^{{N^{(K \!+ \! {K_{min}})}}}} \right):~\beta _{{N^{({K_{min}})}}}^{{N^{(K \!+ \! {K_{min}})}}} \le \rho } I\left( {\left. {{\bf{x}}_{{N^{({K_{min}})}}}^{{N^{(K \!+ \! {K_{min}})}}};{\bf{r}}_{{N^{({K_{min}})}}}^{{N^{(K \!+ \! {K_{min}})}}}} \right|\left\{ {{\bf{x}}_{{N^{({K_{min}})}} \! - \! L \!+ \! 1}^{{N^{({K_{min}})}}},{\bf{w}}_{{N^{({K_{min}})}} \! - \! L \!+ \! 1}^{{N^{({K_{min}})}}}} \right\} \! = \! {{\bf{s}}_0}} \right) \notag \\
&\quad\stackrel{(e)}{\! = \!} \mathop { \inf }\limits_{{\bf s}_0 \in \mathcal{S}} \mathop {\sup }\limits_{p\left( {\bf x}^{N^{(K)}} \right):~{\beta_{N^{(K)}}} \le \rho }I\left(\left.{\bf x}^{N^{(K)}};{\bf r}^{N^{(K)}}\right|{\bf S}_0 \! = \! {\bf s}_0\right) \notag \\
&\quad\! = \! \mathop { \inf }\limits_{{\bf s}_0 \in \mathcal{S}}{N^{(K)}}C_{N^{(K)}}\left( {{{{\bf s} }_0}} \right), \label{eqn:proof2a}
\end{align}
where $(a)$ follows since adding power constraints can only reduce the supremum, and from the mutual information chain rule \cite[Ch. 2.4]{Cover:00}, as
\begin{align*}
I\left( {{{\bf{x}}^{{N^{\left( {K \!+ \! {K_{min}}} \right)}}}};{\bf{r}}_{L - 1}^{{N^{\left( {K \!+ \! {K_{min}}} \right)}}}} \right)
&= I\left( {{{\bf{x}}^{{N^{\left( {K \!+ \! {K_{min}}} \right)}}}};{\bf{r}}_{L - 1}^{{N^{\left( {{K_{min}}} \right)}}}} \right) \!+ \! I\left( {\left. {{{\bf{x}}^{{N^{\left( {K \!+ \! {K_{min}}} \right)}}}};{\bf{r}}_{{N^{\left( {{K_{min}}} \right)}}}^{{N^{\left( {K \!+ \! {K_{min}}} \right)}}}} \right|{\bf{r}}_{L - 1}^{{N^{\left( {{K_{min}}} \right)}}}} \right) \\
&= I\left( {{{\bf{x}}^{{N^{\left( {K \!+ \! {K_{min}}} \right)}}}};{\bf{r}}_{L - 1}^{{N^{\left( {{K_{min}}} \right)}}}} \right) \!+ \! I\left( {\left. {{{\bf{x}}^{{N^{\left( {{K_{min}}} \right)}}}};{\bf{r}}_{{N^{\left( {{K_{min}}} \right)}}}^{{N^{\left( {K \!+ \! {K_{min}}} \right)}}}} \right|{\bf{r}}_{L - 1}^{{N^{\left( {{K_{min}}} \right)}}}} \right) \\
& \qquad \quad  \!+ \! I\left( {\left. {{\bf{x}}_{{N^{\left( {{K_{min}}} \right)}}}^{{N^{\left( {K \!+ \! {K_{min}}} \right)}}};{\bf{r}}_{{N^{\left( {{K_{min}}} \right)}}}^{{N^{\left( {K \!+ \! {K_{min}}} \right)}}}} \right|{{\bf{x}}^{{N^{\left( {{K_{min}}} \right)}}}},{\bf{r}}_{L - 1}^{{N^{\left( {{K_{min}}} \right)}}}} \right) \\
&\geq I\left( {\left. {{\bf{x}}_{{N^{\left( {{K_{min}}} \right)}}}^{{N^{\left( {K \!+ \! {K_{min}}} \right)}}};{\bf{r}}_{{N^{\left( {{K_{min}}} \right)}}}^{{N^{\left( {K \!+ \! {K_{min}}} \right)}}}} \right|{{\bf{x}}^{{N^{\left( {{K_{min}}} \right)}}}},{\bf{r}}_{L - 1}^{{N^{\left( {{K_{min}}} \right)}}}} \right);
\end{align*}
$(b)$ follows from the definition of the LPGC \eqref{eqn:model_LPGC1}, as the the input-output relationship is affected only on the previous $L-1$ channel inputs and channel outputs, and ${\bf{w}}_{{N^{\left( {{K_{min}}} \right)}} - L + 1}^{{N^{\left( {{K_{min}}} \right)}}}$ can be deterministically obtained from ${{\bf{x}}^{{N^{\left( {{K_{min}}} \right)}}}}$ and ${\bf{r}}_{L - 1}^{{N^{\left( {{K_{min}}} \right)}}}$;
$(c)$ follows from the definition of the conditional mutual information \cite[Ch. 2.4]{Gallager:68}, noting that $\tilde{\bf s}_0$ represents the realization of $L-1$ pairs of samples of $x[n]$ and $w[n]$;
To justify $(d)$, we first define the mutual information evaluated with a PDF $\tilde{p}\left(\cdot\right)$ on the channel inputs ${\bf{x}}_{{N^{\left( {{K_{min}}} \right)}}}^{{N^{\left( {K + {K_{min}}} \right)}}}$ as
\begin{equation*}
 I\bigg(\! \left. {\bf{x}}_{{N^{\left( {{K_{min}}} \right)}}}^{{N^{\left( {K \! + \! {K_{min}}} \right)}}};{\bf{r}}_{{N^{\left( {{K_{min}}} \right)}}}^{{N^{\left( {K \! + \! {K_{min}}} \right)}}} \right|\left\{ {{\bf{x}}_{{N^{\left( {{K_{min}}} \right)}} \! - \! L \! + \! 1}^{{N^{\left( {{K_{min}}} \right)}}},{\bf{w}}_{{N^{\left( {{K_{min}}} \right)}} \! - \! L \! + \! 1}^{{N^{\left( {{K_{min}}} \right)}}}} \right\} \! = \! {{\bf{s}}_0} \bigg)_{\tilde p}.
\end{equation*}
Since the conditional mutual information is continuous, it follows from the mean value theorem for integration \cite[Ch. 10.2]{Malik:92}
that for any $\tilde{p}\left(\cdot\right)$, $\exists {\bf s}'_0 \in \mathcal{S}$ such that
\begin{align}
&\int\limits_{{{\bf{\tilde s}}_0} \in \mathcal{S}}\!\! {I\left( {\left. {{\bf{x}}_{{N^{\left( {{K_{min}}} \right)}}}^{{N^{\left( {K \!+ \! {K_{min}}} \right)}}};{\bf{r}}_{{N^{\left( {{K_{min}}} \right)}}}^{{N^{\left( {K \!+ \! {K_{min}}} \right)}}}} \right|\left\{ {{\bf{x}}_{{N^{\left( {{K_{min}}} \right)}} \! - \! L \!+ \! 1}^{{N^{\left( {{K_{min}}} \right)}}},{\bf{w}}_{{N^{\left( {{K_{min}}} \right)}} \! - \! L \!+ \! 1}^{{N^{\left( {{K_{min}}} \right)}}}} \right\} \!= \! {{\bf{\tilde  s}}_0}} \right)_{\tilde{p}}} {p_{{{\bf{x}}_{{N^{\left( {{K_{min}}} \right)}} \! - \! L \!+ \! 1}^{{N^{\left( {{K_{min}}} \right)}}},{\bf{w}}_{{N^{\left( {{K_{min}}} \right)}} \! - \! L \!+ \! 1}^{{N^{\left( {{K_{min}}} \right)}}}}}}\!\!\!\!\!\!\left( {{{\bf{\tilde  s}}_0}} \right)d{{\bf{\tilde s}}_0} \notag \\
& \qquad \!= \! {I\left( {\left. {{\bf{x}}_{{N^{\left( {{K_{min}}} \right)}}}^{{N^{\left( {K \!+ \! {K_{min}}} \right)}}};{\bf{r}}_{{N^{\left( {{K_{min}}} \right)}}}^{{N^{\left( {K \!+ \! {K_{min}}} \right)}}}} \right|\left\{ {{\bf{x}}_{{N^{\left( {{K_{min}}} \right)}} \! - \! L \!+ \! 1}^{{N^{\left( {{K_{min}}} \right)}}},{\bf{w}}_{{N^{\left( {{K_{min}}} \right)}} \! - \! L \!+ \! 1}^{{N^{\left( {{K_{min}}} \right)}}}} \right\} \!= \! {{\bf{s}}'_0}} \right)_{\tilde{p}}} \notag \\
& \qquad \qquad \qquad \times \int\limits_{{{\bf{\tilde s}}_0} \in \mathcal{S}} {p_{{{\bf{x}}_{{N^{\left( {{K_{min}}} \right)}} - L + 1}^{{N^{\left( {{K_{min}}} \right)}}},{\bf{w}}_{{N^{\left( {{K_{min}}} \right)}} - L + 1}^{{N^{\left( {{K_{min}}} \right)}}}}}}\left( {{{\bf{\tilde  s}}_0}} \right)d{{\bf{\tilde s}}_0} \notag \\
& \qquad = I\left( {\left. {{\bf{x}}_{{N^{\left( {{K_{min}}} \right)}}}^{{N^{\left( {K + {K_{min}}} \right)}}};{\bf{r}}_{{N^{\left( {{K_{min}}} \right)}}}^{{N^{\left( {K + {K_{min}}} \right)}}}} \right|\left\{ {{\bf{x}}_{{N^{\left( {{K_{min}}} \right)}} - L + 1}^{{N^{\left( {{K_{min}}} \right)}}},{\bf{w}}_{{N^{\left( {{K_{min}}} \right)}} - L + 1}^{{N^{\left( {{K_{min}}} \right)}}}} \right\} = {{\bf{s}}'_0}} \right)_{\tilde{p}}. \label{eqn:MVT2}
\end{align}
Note that the mean value theorem for integration requires the integral to be defined over a finite interval. However, since
\begin{equation*}
\int\limits_{{{\bf{\tilde s}}_0} \in \mathcal{S}} {p_{{{\bf{x}}_{{N^{\left( {{K_{min}}} \right)}} - L + 1}^{{N^{\left( {{K_{min}}} \right)}}},{\bf{w}}_{{N^{\left( {{K_{min}}} \right)}} - L + 1}^{{N^{\left( {{K_{min}}} \right)}}}}}}\left( {{{\bf{\tilde  s}}_0}} \right)d{\bf{\tilde s}}_0 = 1,
\end{equation*}
it follows that the probability density function approaches 0 for $\left\|{{{\bf{\tilde  s}}_0}}\right\| \rightarrow \infty$, therefore the integral can be approached arbitrarily close by considering finite intervals, i.e., integrating over $\left\|{{{\bf{\tilde  s}}_0}}\right\| \leq \Omega$ for sufficiently large $\Omega$, instead of over $\mathcal{S} = \mathds{R}^{2L}$.
Let $p^*\left(\cdot\right)$ denote the input PDF which maximizes
\begin{equation*}
\mathop {\sup }\limits_{p\left( {{\bf{x}}_{{N^{({K_{min}})}}}^{{N^{(K \! + \! {K_{min}})}}}} \right):~\beta _{{N^{({K_{min}})}}}^{{N^{(K \! + \! {K_{min}})}}} \le \rho } I\left( {\left. {{\bf{x}}_{{N^{({K_{min}})}}}^{{N^{(K \! + \! {K_{min}})}}};{\bf{r}}_{{N^{({K_{min}})}}}^{{N^{(K \! + \! {K_{min}})}}}} \right|\left\{ {{\bf{x}}_{{N^{({K_{min}})}} \! - \! L \! + \! 1}^{{N^{({K_{min}})}}},{\bf{w}}_{{N^{({K_{min}})}} \! - \! L \! + \! 1}^{{N^{({K_{min}})}}}} \right\} \! = \! {{\bf{s}}_0}} \right).
\end{equation*}
It then follows that
\begin{align*}
&\mathop {\inf }\limits_{{{\bf{s}}_0} \in \mathcal{S}} \mathop {\sup }\limits_{p\left( {{\bf{x}}_{{N^{({K_{min}})}}}^{{N^{(K \! + \! {K_{min}})}}}} \right):~\beta _{{N^{({K_{min}})}}}^{{N^{(K \! + \! {K_{min}})}}} \le \rho } I\left( {\left. {{\bf{x}}_{{N^{({K_{min}})}}}^{{N^{(K \! + \! {K_{min}})}}};{\bf{r}}_{{N^{({K_{min}})}}}^{{N^{(K \! + \! {K_{min}})}}}} \right|\left\{ {{\bf{x}}_{{N^{({K_{min}})}} \! - \! L \! + \! 1}^{{N^{({K_{min}})}}},{\bf{w}}_{{N^{({K_{min}})}} \! - \! L \! + \! 1}^{{N^{({K_{min}})}}}} \right\} \! = \! {{\bf{s}}_0}} \right) \\
& = \! \mathop {\inf }\limits_{{{\bf{s}}_0} \in \mathcal{S}}  I\left( {\left. {{\bf{x}}_{{N^{({K_{min}})}}}^{{N^{(K \! + \! {K_{min}})}}};{\bf{r}}_{{N^{({K_{min}})}}}^{{N^{(K \! + \! {K_{min}})}}}} \right|\left\{ {{\bf{x}}_{{N^{({K_{min}})}} \! - \! L \! + \! 1}^{{N^{({K_{min}})}}},{\bf{w}}_{{N^{({K_{min}})}} \! - \! L \! + \! 1}^{{N^{({K_{min}})}}}} \right\} \! = \! {{\bf{s}}_0}} \right)_{p^*} \\
&\stackrel{(i)}{\leq}  I\left( {\left. {{\bf{x}}_{{N^{\left( {{K_{min}}} \right)}}}^{{N^{\left( {K \! + \! {K_{min}}} \right)}}};{\bf{r}}_{{N^{\left( {{K_{min}}} \right)}}}^{{N^{\left( {K \! + \! {K_{min}}} \right)}}}} \right|\left\{ {{\bf{x}}_{{N^{\left( {{K_{min}}} \right)}} \! - \! L \! + \! 1}^{{N^{\left( {{K_{min}}} \right)}}},{\bf{w}}_{{N^{\left( {{K_{min}}} \right)}} \! - \! L \! + \! 1}^{{N^{\left( {{K_{min}}} \right)}}}} \right\} \! = \! {{\bf{s}}'_0}} \right)_{p^*} \\
&\stackrel{(j)}{\! = \!} \int\limits_{{{\bf{\tilde s}}_0} \in \mathcal{S}} \!\!{I\left( {\left. {{\bf{x}}_{{N^{\left( {{K_{min}}} \right)}}}^{{N^{\left( {K \! + \! {K_{min}}} \right)}}};{\bf{r}}_{{N^{\left( {{K_{min}}} \right)}}}^{{N^{\left( {K \! + \! {K_{min}}} \right)}}}} \right|\left\{ {{\bf{x}}_{{N^{\left( {{K_{min}}} \right)}} \! - \! L \! + \! 1}^{{N^{\left( {{K_{min}}} \right)}}},{\bf{w}}_{{N^{\left( {{K_{min}}} \right)}} \! - \! L \! + \! 1}^{{N^{\left( {{K_{min}}} \right)}}}} \right\} \! = \! {{\bf{\tilde  s}}_0}} \right)_{p^*}}  {p_{{{\bf{x}}_{{N^{\left( {{K_{min}}} \right)}} \! - \! L \! + \! 1}^{{N^{\left( {{K_{min}}} \right)}}},{\bf{w}}_{{N^{\left( {{K_{min}}} \right)}} \! - \! L \! + \! 1}^{{N^{\left( {{K_{min}}} \right)}}}}}}\!\!\!\!\left( {{{\bf{\tilde  s}}_0}} \right)d{{\bf{\tilde s}}_0} \\
& \leq\!\!\!\!\! \mathop {\sup }\limits_{p\left( {{{\bf{x}}^{{N^{\left( {K \! + \! {K_{min}}} \right)}}}}} \right):  ~\beta _{{N^{\left( {K_{min}} \right)}}}^{{N^{\left( {K \! + \! {K_{min}}} \right)}}} \le \rho ; ~{\beta _{{N^{\left( {K_{min}} \right)}}}} \le \rho} ~\int\limits_{{{\bf{\tilde s}}_0} \in \mathcal{S}} {\!\!\!I\left( {\left. {{\bf{x}}_{{N^{\left( {{K_{min}}} \right)}}}^{{N^{\left( {K \! + \! {K_{min}}} \right)}}};{\bf{r}}_{{N^{\left( {{K_{min}}} \right)}}}^{{N^{\left( {K \! + \! {K_{min}}} \right)}}}} \right|\left\{ {{\bf{x}}_{{N^{\left( {{K_{min}}} \right)}} \! - \! L \! + \! 1}^{{N^{\left( {{K_{min}}} \right)}}},{\bf{w}}_{{N^{\left( {{K_{min}}} \right)}} \! - \! L \! + \! 1}^{{N^{\left( {{K_{min}}} \right)}}}} \right\}\! \! = \! \!{{\bf{\tilde  s}}_0}} \right)} \\
& \quad \qquad \qquad \qquad  \qquad \qquad   \qquad \qquad  \qquad \qquad \quad   \times {p_{{{\bf{x}}_{{N^{\left( {{K_{min}}} \right)}} \! - \! L \! + \! 1}^{{N^{\left( {{K_{min}}} \right)}}},{\bf{w}}_{{N^{\left( {{K_{min}}} \right)}} \! - \! L \! + \! 1}^{{N^{\left( {{K_{min}}} \right)}}}}}}\left( {{{\bf{\tilde  s}}_0}} \right)d{{\bf{\tilde s}}_0},
\end{align*}
in $(i)$, ${\bf s}'_0$ is selected such that the mean value theorem as restated in \eqref{eqn:MVT2} is satisfied for some $p\left({\bf x}^{N^{(K_{min})}}\right)$ which satisfies the power constraint $\beta_{N^{(K_{min})}} \leq \rho$; $(j)$ follows from the mean value theorem and the selection of $p\left({\bf x}^{N^{(K_{min})}}\right)$ and  ${\bf s}'_0$ as stated in step $(i)$;
%
$(e)$ follows from the definition of the LPGC \eqref{eqn:model_LPGC1}, as the joint probability function is invariant to index shifting by a number of samples which is an integer multiple of $N_{lcm}$, noting that the RV ${\bf S}_0$ represents $L-1$ pairs of samples of $x[n]$ and $w[n]$.

Dividing both sides of \eqref{eqn:proof2a} by ${N^{(K)}}$ leads to
\begin{align*}
\mathop { \inf }\limits_{{\bf s}_0 \in \mathcal{S}} C_{N^{(K)}}\left( {{{{\bf s} }_0}} \right)
&\leq \frac{1}{N^{(K)}}\mathop {\sup }\limits_{p\left( {\bf x}^{N^{(K+K_{min})}} \right):~{\beta_{N^{(K+K_{min})}}} \le \rho }I\left({\bf x}^{N^{(K+K_{min})}};{\bf r}_{L-1}^{N^{(K+K_{min})}}\right) \\
&= \frac{K+K_{min}}{K}R_{{K+K_{min}}}.
\end{align*}
This proves the lemma.
$\blacksquare$
\smallskip

\begin{lemma}
\label{lem:proof3}
$\mathop {\sup }\limits_{K > K_{min}} R_K \le C_{LPTV-ACGN}$.
\end{lemma}
{\em Proof:}
We now show that for any rate $R$ achievable for a $N^{(K)}$-MPGC, $K > K_{min}$, is also achievable for the LPGC, i.e., for all $\epsilon, \gamma > 0$, if we take a sufficiently large $n$, then there exists a $[R_1, n]$ code for the LPGC such that \eqref{eqn:def_Rs1}-\eqref{eqn:def_Rs3} are satisfied. The proof follows the same guidelines as that of \cite[Lemma 2]{Goldsmith:01}.
We first show this for $n$ an integer multiple of $N^{(K)}$, then we prove this true for all sufficiently large $n$.
Fix $K > K_{min}$ and consider a rate $R$ achievable for the $N^{(K)}$-MPGC. Then, for any $\epsilon, \gamma > 0$, $\exists b_0$ sufficiently large, such that for all integer $b > b_0$, there exists a $[R_1, b\cdot N^{(K)}]$ code for the $N^{(K)}$-MPGC with average error probability which satisfies \eqref{eqn:def_Rs1}, and code rate which satisfies
\begin{equation}
\label{eqn:lem2Proof2b}
R_1  \ge R - \frac{\gamma }{2}.
\end{equation}
We denote this code by $\mathcal{C} _{b \cdot N^{(K)}}$. Note that as the channel is $N^{(K)}$-MPGC, the $\mathcal{C} _{b \cdot N^{(K)}}$ code only considers the last $M^{(K)}$ channel outputs out of each $N^{(K)}$ channel outputs for decoding.

Next, apply the code  $\mathcal{C} _{b \cdot N^{(K)}}$  to the LPGC. The rate is unchanged. Since the decoder considers the last $M^{(K)}$ channel outputs out of each $N^{(K)}$ channel outputs, the error probability is also the same as that of the $N^{(K)}$-MPGC. Consequently, for $n = b \cdot N^{(K)}$, there exists a code for the LPGC with the desired rate and an arbitrarily small error probability for all ${\bf s}_0$, assuming large enough $b$.

We now extend this coding scheme for the LPGC to arbitrary values of $n$. Specifically, we write $n = b \cdot N^{(K)} + a$, where $b$ is an integer and $a \in \mathcal{N}^{(K)}$. We define a $\left[R_1 \frac{{b \cdot N^{(K)}}}{{b \cdot N^{(K)} + a}}, b\cdot N^{(K)} + a\right]$ code for the LPGC by appending $a$ arbitrary symbols to the codewords of $\mathcal{C} _{b \cdot N^{(K)}}$. The decoder discards the last $a$ channel outputs. Clearly, for all ${\bf s}_0$, the error probability is the same as that of the  $\mathcal{C} _{b \cdot N^{(K)}}$ code since the decoders operate on the same received symbols. The code rate is obtained by
\begin{equation*}
R_1 \frac{{b \cdot N^{(K)}}}{{b \cdot N^{(K)} + a}}
\stackrel{(a)}{\ge} \left( {R - \frac{\gamma  }{2}} \right)\frac{{b \cdot N^{(K)}}}{{b \cdot N^{(K)} + a}},
\end{equation*}
where $(a)$ follows from \eqref{eqn:lem2Proof2b}. Thus, for sufficiently large $b$, it follows that
\begin{equation*}
R_1  \ge R - \gamma .
\end{equation*}
We therefore conclude that any rate $R$ achievable for a $N^{(K)}$-MPGC can be obtained by a $[R_1, n]$ code for the LPGC for any sufficiently large $n$.
$\blacksquare$
\smallskip

We conclude that for every $\epsilon \in (0,1)$, $\gamma > 0$, $\exists K_0 >0$ such that $\forall K > K_0$
\begin{align}
\mathop {\sup }\limits_{K > K_{min}} R_K
&\stackrel{(a)}{\leq} C_{LPTV-ACGN} \notag \\
&\stackrel{(b)}{\leq} \frac{1}{1 - {\epsilon }}\mathop {\inf }\limits_{{{{\bf s} }_0} \in \mathcal{S}}C_{N^{(K)}}\left({\bf s}_0\right) + \frac{1}{\left({1 - {\epsilon }}\right){N^{(K)}}} + \gamma \notag  \\
&\stackrel{(c)}{\leq} \frac{K+K_{min}}{\left({1 - {\epsilon }}\right)K}R_{{K+K_{min}}} + \frac{1}{\left({1 - {\epsilon }}\right){N^{(K)}}} + \gamma, \label{eqn:Last1}
\end{align}
 where $(a)$ follows from Lemma \ref{lem:proof3}; $(b)$ follows from Lemma \ref{Cor:Proof_C2}; $(c)$ follows from follows from Lemma \ref{lem:proof2}.
As \eqref{eqn:Last1} is satisfied for all sufficiently large $K$, it follows that
\begin{align}
C_{LPTV-ACGN}
&\leq \mathop {\lim \inf }\limits_{K \to \infty } \left(\frac{K+K_{min}}{\left({1 - {\epsilon }}\right)K}R_{{K+K_{min}}} +  \frac{1}{\left({1 - {\epsilon }}\right){N^{(K)}}}\right) + \gamma \notag \\
&=  \frac{1}{1 - {\epsilon }}\mathop {\lim \inf }\limits_{K \to \infty } R_{K}  + \gamma. \label{eqn:Last2b}
\end{align}
Since  $\epsilon , \gamma$ can be made arbitrarily small, \eqref{eqn:Last2b} implies that
\begin{equation}
\label{eqn:Last3}
C_{LPTV-ACGN} \leq \mathop {\lim \inf }\limits_{K \to \infty } R_K.
\end{equation}
Lastly, it follows from the definition of $\lim \sup$ \cite[Def. 5.4]{Amann:05} that $\mathop {\lim \sup }\limits_{K \to \infty }R_K \leq \mathop {\sup }\limits_{K > K_{min}} R_K$.
Since $\mathop {\lim \inf }\limits_{K \to \infty } R_K \leq \mathop {\lim \sup }\limits_{K \to \infty }R_K$, it follows that \eqref{eqn:Last1} yields
\begin{equation*}
C = \mathop {\lim }\limits_{K \to \infty }R_K.
\end{equation*}
We have therefore shown that the capacity of the discrete time LPTV channel with ACGN can be obtained, to arbitrary accuracy, from the capacity of the equivalent channel with a finite number of of channel outputs discarded on each block, where the size of each block is an integer multiple of both the period of the LPTV channel and the period of the cyclostationary noise, for all sufficiently large block sizes.


\vspace{-0.2cm}
\chapter{Proof of Thm. \ref{thm:CapacityLPTV2}}
\label{app:Proof2}
\vspace{-0.1cm}
The result of Thm. \ref{thm:CapacityLPTV2} is based on the capacity of finite memory multivariate Gaussian channels derived in \cite{Brandenburg:74}.
Define the $N_0 \times 1$ vectors $\tilde{\bf x}[n]$ and $\tilde{\bf r}[n]$  whose elements are given by  $\left(\tilde{\bf x}[n]\right)_i = x\left[nN_0 + i\right]$ and $\left(\tilde{\bf r}[n]\right)_i = r\left[nN_0 + i\right]$, respectively, $i \in {\mathcal{N}}_0$.
Recalling the definitions of $\tilde{\bf w}[n]$ and ${\bf H}[l]$ for $l=0,1$, we note that \eqref{eqn:ChannelModel1} can be transformed into
\begin{equation}
\label{eqn:FDModel1}
\tilde{\bf r}[n] = \sum \limits_{l=0}^1 {\bf H}[l]\tilde{\bf x}[n-l] + \tilde{\bf w}[n].
\end{equation}
Since  $\tilde{\bf w}[n]$ denotes the DCD of the ACGN $w[n]$, it follows that $\tilde{\bf w}[n]$ is a stationary multivariate Gaussian stochastic process.
Moreover, as the statistical dependance of $w[n]$ spans a finite interval, it follows that the statistical dependance of the transformed multivariate noise $\tilde{\bf w}[n]$ also spans a finite interval. Thus, \eqref{eqn:FDModel1} models a multivariate stationary Gaussian channel with finite memory. Let $B$ denote the codeword length in the equivalent MIMO channel, since the transmitted vector is subject to power constraint \eqref{eqn:Cap_Constraint},  the transmitted signal in the equivalent channel \eqref{eqn:FDModel1} is subject to the  average power constraint
\begin{equation}
\label{eqn:FDModelConst}
\frac{1}{B \cdot N_0}\sum\limits_{n=0}^{B-1}\sum\limits_{k=0}^{N_0-1}\E\left\{\left|\left(\tilde{\bf x}[n]\right)_k\right|^2\right\} \leq \rho.
\end{equation}
Note that the scalar channel output is obtained from $\tilde{\bf r}[n]$ using the inverse DCD, hence the transformation which obtains \eqref{eqn:FDModel1} from \eqref{eqn:ChannelModel1} is reversible, and the capacity of the equivalent channel \eqref{eqn:FDModel1} is clearly equal to the capacity of the LPTV channel with ACGN \eqref{eqn:ChannelModel1}.
Following the capacity derivation of multivariate stationary Gaussian channels with memory in \cite{Brandenburg:74}\footnote{We note that \cite[Thm. 1]{Brandenburg:74} is stated for a per-codeword constraint. However, it follows from the proof of \cite[Lemma 3]{Brandenburg:74} and from \cite[Ch. 7.3, pgs. 323-324]{Gallager:68} that \cite[Thm. 1]{Brandenburg:74} holds also with an average power constraint \eqref{eqn:FDModelConst}.}, let $\tilde{\Delta}$ denote the unique solution to \cite[Eqn. (9a)]{Brandenburg:74}\footnote{\cite[Thm. 1]{Brandenburg:74} is stated as follows: {\em Let the $s \times s$ matrix $\Gamma(\theta) = H(\theta)^{-1}R(\theta)H(\theta)^{-*}$, and let $\lambda_1(\theta), \lambda_2(\theta), \ldots, \lambda_s(\theta)$ be the eigenvalues of $\Gamma(\theta)$, $-\pi \leq \theta \leq \pi$. Then, $\lambda_j(\theta) > 0$, $1 \leq j \leq s$, $-\pi \leq \theta \leq \pi$. Let $S \geq 0$ be given, and let $K_s$ be the (unique) positive number such that
\begin{equation*}
\frac{1}{2\pi}\sum\limits_{j=1}^{s}\int\limits_{-\pi}^{\pi}d\theta \max \left[0, K_s - \lambda_j(\theta)\right] = S.
\end{equation*}
Then
\begin{equation*}
C_s = \frac{1}{4\pi}\sum\limits_{j=1}^{s}\int\limits_{-\pi}^{\pi}d\theta \max \left(0, \log_2 \frac{K_s}{\lambda_j(\theta)}\right).
\end{equation*}
 } The power constraint is given by $\frac{1}{N}\sum\limits_{n=0}^{N-1}\left\|x_i(n)\right\|^2 \leq S$ for each message $i$ \cite[Eqn. 3]{Brandenburg:74}, and the $s \times s$ matrices $H(\theta)$ and $R(\theta)$ are obtained as the DTFT of the multivariate CTF \cite[Eqn. 5]{Brandenburg:74} and of the autocorrelation function of the multivariate noise \cite[Eqn. 8]{Brandenburg:74}, respectively.}
\begin{equation*}
\frac{1}{2\pi }\sum\limits_{k=0}^{N_0 -1} \int\limits_{\omega = -\pi}^{\pi} \left(\tilde{\Delta} - \left(\tilde{\lambda} _k (\omega)\right)^{-1}\right)^+ d\omega = \rho \cdot N_0,
\end{equation*}
then the capacity of \eqref{eqn:FDModel1} is given by \cite[Eqn. (9b)]{Brandenburg:74}
\begin{equation}
\label{eqn:FDCap1}
C_{eq} = \frac{1}{4\pi}\sum\limits_{k=0}^{N_0 -1} \int\limits_{\omega = -\pi}^{\pi} \left(\log \left({\tilde{\Delta}}\cdot{\tilde{\lambda} _k (\omega)}\right)\right)^+ d\omega.
\end{equation}
Note that \eqref{eqn:FDCap1} is in units of bits per MIMO channel use, as it denotes the capacity of the equivalent MIMO channel. Since each MIMO channel use corresponds to $N_0$ scalar channel uses, the capacity of the original scalar channel is thus
\begin{equation*}
C_{LPTV-ACGN}  = \frac{1}{4\pi N_0}\sum\limits_{k=0}^{N_0 -1} \int\limits_{\omega = -\pi}^{\pi} \left(\log \left({\tilde{\Delta}} \cdot {\tilde{\lambda} _k (\omega)}\right)\right)^+ d\omega.
\end{equation*}	
Lastly, note that the capacity achieving $\tilde{\bf x}[n]$ is a zero-mean  $N_0 \times 1$ multivariate stationary Gaussian process \cite{Brandenburg:74}, thus, the for the scalar channel, the capacity achieving $x[n]$, obtained via the inverse DCD of $\tilde{\bf x}[n]$, is a cyclostationary Gaussian process with period $N_0$.
This completes the proof. 
$\blacksquare$

\end{appendix}

\end{document}